\documentclass[twocolumn]{aastex631}

\usepackage{graphicx}	% Including figure files
\usepackage{amsmath,esint}
\usepackage{url}
\usepackage{natbib}
\usepackage{multirow}
\usepackage{bm}
\usepackage{xspace}
\usepackage{xcolor}
\usepackage[utf8]{inputenc}
\usepackage{hyperref}
\usepackage{booktabs}
\usepackage{longtable}
\usepackage{tabularx}

\newcommand{\asat}{{\em AstroSat}}
\newcommand{\fermi}{{\em Fermi}}
\newcommand{\swift}{{\em Swift}}
\newcommand{\czti}{{\asat-CZTI}}

\newcommand{\gfo}{GW170817}

\newcommand{\egs}{$\mathrm{ergs~cm^{-2}~s^{-1}}$}
\newcommand{\eg}{$\mathrm{ergs~s^{-1}}$}

\newcommand{\msun}{\ensuremath{\mathrm{M}_\odot}}

\begin{document}

\title{Bright in the Black: Searching for Electromagnetic Counterparts to Gravitational-Wave Candidates in LIGO-Virgo-KAGRA Observation Runs with \czti}

\author[0000-0003-3630-9440]{Gaurav Waratkar}
\affiliation{Department of Physics, Indian Institute of Technology Bombay, Powai, 400 076, India}

\author[0000-0002-6112-7609]{Varun Bhalerao}
\affiliation{Department of Physics, Indian Institute of Technology Bombay, Powai, 400 076, India}

\author[0000-0003-3352-3142]{Dipankar Bhattacharya}
\affiliation{Ashoka University, Department of Physics, Sonepat, Haryana 131029, India}
\affiliation{Inter-University Center for Astronomy and Astrophysics, Pune, Maharashtra-411007, India}

% \author{Friends}
% \affiliation{\czti}

\correspondingauthor{G. Waratkar}
\email{gauravwaratkar@iitb.ac.in}

\begin{abstract}
	GW150914 marked the start of the gravitational wave (GW) era with the direct detection of binary black hole (BBH) merger by the LIGO-Virgo GW detectors. The event was temporally coincident with a weak signal detected by \fermi-GBM, which hinted towards the possibility of electromagnetic emission associated with the compact object coalescence. The detection of a short Gamma-Ray Burst (GRB) associated with GW1708017, along with several multi-wavelength detections, truly established that compact object mergers are indeed multi-messenger events. The Cadmium Zinc Telluride Imager onboard \asat\ can search for X-ray counterparts of the GW events and has detected over 600 GRBs since launch. Here we present results from our searches for counterparts coincident with GW triggers from the first three LIGO-Virgo-KAGRA (LVK) GW Transient Catalogs. For 72 out of 90 GW events for which \czti\ data was available, we undertook a systematic search for temporally coincident transients and we detected no X-ray counterparts. We evaluate the upper limits on the maximum possible flux from the source in a 100~s window centered around each trigger, consistent with the GW localization of the event. Thanks to the high sensitivity of CZTI, these upper limits are highly competitive with those from other spacecraft. We use these upper limits to constrain the theoretical models that predict high-energy counterparts to the BBH mergers. We also discuss the probability of non-detections of BBH mergers at different luminosities and the implications of such non-detections from the ongoing fourth observing run of the LVK detectors.

\end{abstract}

% Nobody found anything. We found nothing. Yet we all are here. Aren't we all salesmen in a way? Where's the lie?

%%%%%%%%%%%%%%%%%%%%%%%%%%%%%%%%%%%%%%%%%%%%%%%%%%

\section{Introduction} \label{sec:intro}

In the year 2015, LIGO-Virgo gravitational wave (GW) detectors reported the first direct detection of gravitational waves from a binary black hole (BBH) merger \citep{gw150914lvk2016}. This event was accompanied by the detection of a weak X-ray transient signal in \fermi-GBM \citep[GRB150914-GBM; ][]{connaughton2016, veres2019}, which was not seen by any other mission. This unproven association of the electromagnetic (EM) signal with the GW event started the multi-messenger era. This era truly reached its peak when, in 2017, LIGO-Virgo GW detectors reported the first detection of GW from a binary neutron star merger (BNS) in 2017. About 1.7~s after the detection of GW, a short-duration gamma-ray burst (GRB) was detected by INTEGRAL, \fermi\ \citep{grbgw2017}. Soon, the counterpart to this GW event was detected across the electromagnetic spectrum in the form of a kilonova (KN) emission \citep{bnsmma2017}. To date, this remains the only known multi-messenger event where gravitational waves from the merger, prompt emission from GRB, and a corresponding KN emission were all detected \citep{petrovdatadriven2022}. It further established that compact object mergers like BNS mergers are one of the progenitors for GRBs that typically last for less than 2~s \citep{fermi1708172017, bergersgrbreview2014, metzgerberger2013}.

Other compact object mergers that the GW detectors can detect include the merger of a neutron star and black hole \citep[NSBH;][]{twonsbhLVK2021, bv190814LVK2020, lvkprospects2018}. These NSBH mergers can also be accompanied by EM emission depending on the mass ratios of the system, surrounding environment, and so on \citep{pirannakarCOMMA2013, akshat2022, nakar2020}. In addition, GW detectors have been regularly detecting binary black hole (BBH) mergers \citep{gw150914lvk2016, gwtc1-2019, gwtc2-2021, gwtc3-2021}. While most studies point towards the need for at least one neutron star in the binary system to get an EM emission, some models suggest that even BBH mergers can lead to EM emissions \citep{mckernanGRBBBH2019, pernaEMinAGN2021, bartosBBHorigin}.

In the past three observing runs of LIGO-Virgo GW detectors, there have been several attempts to hunt for EM emission coming from these sources, none of which led to any proven counterpart to the GW events \citep{petrovdatadriven2022}. The non-detections from campaigns that hunted for the afterglow emission were used to improve the models that predict the EM emission \citep[see for instance][]{harsh26c2022, shreyaNat2021, growth25z2019, growthbv2020}. \citet{kasliwalluminosity2020} used the optical non-detections to constrain the KN luminosity function, and place constraints on bright-KN rates. Recent studies have also discussed the association between Fast Radio Bursts (FRBs) and GW events, although none conclusive yet \citep{frbgwNAT2023, mohitFRBGW2023, frbgwO3a2022}. Furthermore, the large distances at which most of the BBHs are detected and the challenges associated with the formation of accretion disks during their merger evolution make it difficult to detect any EM signatures from them. Yet, there have been several claimed associations --- for instance, there were several candidate counterparts to BBH mergers during the third observing run (O3) that were associated with active galactic nuclei (AGNs) with unusual flaring activity \citep{grahamonecanddiate2020, graham2023}.

In the hunt for coincident prompt high-energy emission from sources of GWs, there were non-detections reported in the burst searches reported by various space telescopes like \fermi-GBM, \swift-BAT, {\em AGILE}, {\em Insight}-HXMT, etc \citep{fermiO1search2019, jointgrbgwO1O22020, jointlvkgrb2023, agileGWTC12022, insightGRBGW2021}. Offline searches undertaken to look for coincident events with the GRB events detected by \fermi\ and \swift-BAT during O3 \citep{gwcrossgrbO3a2021, gwcrossgrbO3b2022} also yielded non-detections. 
The \fermi-GBM detected gamma-ray signal coincident with the first BBH merger event GW150914~\citep{connaughton2016} remains the only claimed association between the BBH merger event and the detection of a GRB. Through this detection, \citet{veres2019} calculated the rates of such coincident signals given the sensitivity of LIGO-Virgo GW detectors.

Through these studies, it was seen that both coincident searches and the hunt for the afterglows can vastly improve our understanding of these sources. Such searches will be of even higher importance when the sensitivities of the GW detectors improve in the future observing runs leading to a larger number of GW events \citep{lvkprospects2018}. However, \citet{petrovdatadriven2022} showed that the well-localised, high-profile events will make up a smaller proportion of the overall population that LIGO-Virgo-KAGRA (LVK) GW detectors will be detecting the future observing runs because of improvements in data analysis. Their data-driven approach also pointed out that while the number of GW events for which KN signatures are detectable increases, there is a significant reduction in their fraction among total GW events. This combined with a large number of events, most having poor localisation, underscores the need to have X-ray coincident searches guiding these searches, similar to the case of \gfo\ where the GRB detection set the tone for further follow-up. We see the need to undertake searches for bursts that are coincident with the GW events, which will help us understand the source properties, and also aid in the searches for KNe by other facilities.

The Indian space telescope \asat~\citep{astrosat2014} was launched in September 2015, soon after the discovery of GW150914. One of the instruments on board is the Cadmium Zinc Telluride Imager \citep[CZTI;][]{bhalerao2017} --- a wide-field hard X-ray monitor operating in the 20-200~keV energy range. At energies above $\sim100$~keV, the collimators of CZTI become increasingly transparent to radiation, and CZTI functions as an all-sky monitor with good sensitivity to transients \citep{cift2021}. CZTI has detected over 600 Gamma Ray Bursts (GRBs) to date\footnote{CZTI GRB detections are reported at \url{http://astrosat.iucaa.in/czti/?q=grb}.}, of which $\sim$~50 have not been detected by any other mission. This demonstrates the sensitivity of CZTI to high-energy transients and makes it an important asset in the search for EM counterparts to GW sources.

Since launch, \czti\ is used to regularly search for these GW events based on the low-latency information available immediately to the astronomers, and has placed upper-limits on fluxes from the GW events \citep[see for instance][]{2017GCN.20648....1M,2023GCN.34327....1W}. Further, for the case of GW170817, the primary localisation region was occulted by the earth. Based on the reported source spectrum and flux \citep{2017GCN.21520....1V}, it was inferred that \czti\ was sensitive enough to have detected the short GRB from the event, but it did not. This non-detection was used in constraining the combined localisation region of GW event and the \fermi-GBM GRB detection localisation region \citep{kasliwalscience2017}, highlighting how non-detections can be of immense importance.

With an improved CZTI data processing pipeline \citep{ratheesh2021}, and availability of final confident events published by LVK through the Gravitational Wave Transient Catalogs (GWTC), we report our comprehensive search for bursts in \czti\ coincident with the GW events reported in these catalogs from the first three observing runs (O1, O2, and O3; hereafter referred as O1-3) of LIGO-Virgo GW detectors. In Section~\ref{sec:dataproc}, we describe the procedure for CZTI transient analysis, focusing on the searches for GW counterparts in CZTI data. We outline the GW event sample used, and the results from our searches for coincident bursts in CZTI data in Section~\ref{sec:results}. Finally, we discuss the implications of these results on the luminosity function of different merger models and conclude with the prospects of detecting prompt emission counterparts in the future observing runs of LVK GW detectors in Section~\ref{sec:discussion}.

\section{CZTI Data Processing}\label{sec:dataproc}
The procedure for CZTI transient analysis differs from that of usual persistent sources, and more so for the analysis of GW coincident searches. In this section, we outline the method used for the analysis to search for GW counterparts in CZTI data.

\subsection{Searches for coincident bursts}\label{subsec:czti-pipeline}

We broadly follow the data processing described in \citet{cift2021} and \citet{akash2020}. To reduce the raw data downlinked from the \asat, we use the CZTI V3 pipeline\footnote{\href{http://astrosat-ssc.iucaa.in/cztiData}{http://astrosat-ssc.iucaa.in/cztiData}} which is based on the new generalised event selection algorithm \citep{ratheesh2021}. The unprocessed raw data files (Level 1 data products) are processed through the standard workflow of the CZTI pipeline to get the Level 2 data products, which include the cleaned event data. We use the \texttt{cztgtigen} module to make sure that the data are not discarded even when the on-axis target of \asat~is earth occulted. In the CZTI V3 pipeline, we employ the new ``\texttt{generalized}'' event selection algorithm mode which is optimized for bright GRB-like events. This \texttt{generalized} mode employs new modules \texttt{cztnoisypixclean}, \texttt{cztsuperbunchclean}, \texttt{cztheavybunchclean}, \texttt{cztflickpixclean}, and \texttt{czteventsep} \citep{ratheesh2021}.

Using these cleaned event files, we generate lightcurves that are binned at three timescales --- 0.1~s, 1.0~s, and 10.0~s for all four quadrants. In these lightcurves, we expect a slow variation in the background due to an increased charged particle background when \asat\ is passing over the South Atlantic Anomaly (SAA). We correct this variation by using a second-order Savitzky Golay filter (time window width of 100~s) to estimate the broad background trend, which is then subtracted from the time series data to obtain de-trended light curves. The data are first inspected visually in the form of light curves as well as spectrograms (2-dimensional histograms of photon energies and arrival times), to search for transient candidates. This is followed by a qualitative search for the EM counterpart in the data. 
% In the first step of our search, we visually inspect the spectrogram plots (time vs energy plots), and the time series generated for the search window to search for GRB-like emission signatures for each quadrant.

\subsection{Calculating the minimum detectable flux}\label{subsec:fluxlimits}
For each quadrant, we calculate the minimum counts needed from a source to get a detection at a 90\% confidence level. We then examine the data to see if the light curve crosses these ``cutoff count-rates'' at any point in the search window of 100~s. A valid transient should be detected above the minimum detectable count-rate from any source for each quadrant (cutoff count-rates). We check if anything crosses this threshold, and require that any astrophysical event should be detected in at least three quadrants out of four. If no such candidate is found, we use these count-rates to calculate the upper limits on the source flux.

We follow the same treatment as described in \cite{akash2020} to calculate the cutoff count-rates for each quadrant at three different search timescales --- at 0.1~s, 1.0~s, and 10.0~s. We choose a window of 100~s ($\pm$ 50~s from the trigger time) and use de-trended data from ten neighboring orbits to estimate the background count-rate histograms. We then estimate a cutoff count-rate from these background count-rate distributions based on the figure of merit of false alarm probability (FAP), which is set at 0.1 for each quadrant. The combined four-quadrant FAP becomes $10^{-4}$, which states that we expect an outlier in every $10^4$ bins. The choice of 10~neighbouring orbits allows us to account for temporal background fluctuations of the CZT detectors. It is observed that in some cases, one of the CZTI quadrants is very noisy, and should be excluded from further analysis. We define such ``noisy'' cases as ones where the cutoff count-rate for one quadrant is $>$~3$\sigma$ away from the median of the three quadrants with the lowest cutoff count-rates and remove that quadrant from the further analysis.

The sensitivity of \asat-CZTI varies significantly over the entire sky, and so does the conversion of detector counts to source flux. For every GW source, we calculate direction-dependent flux limits using the \asat\ Mass Model \citep{mate2021}. The satellite response has been pre-calculated by the CZTI team over a 2048-point ``Level 4'' Hierarchical Triangular Mesh (HTM) grid. We downgrade the GW healpix skymap \citep{singerBayestar2016} to \texttt{NSIDE=16} to roughly match this grid resolution, and assign nearest-neighbor sensitivity for each point. We use a simple power-law spectrum $N(E) \sim E^{\Gamma}$, with a $\Gamma = -1$, to convert the count-rates into a flux upper-limit for a given grid point of the Mass Model.  The final flux limits ($F_\mathrm{UL}$) are GW probability-weighted averages over the sky visible to \asat\ at the time of the GW event. We repeat this analysis for all three search timescales.

We estimate the isotropic equivalent luminosities ($L_{iso}$) in the cosmological rest frame for the standard energy range of 1~keV -- 10~MeV through the following equation:

\begin{equation}
	L_\mathrm{iso} = 4 \pi \, D_\mathrm{L}^2 \, F_\mathrm{UL} \, k
\end{equation}
where $k$ is the k-correction factor to correct for the source redshift as well as the energy range of our detectors, defined as follows:

\begin{equation}
	k =
	\frac{ \int_{1 \, \mathrm{keV} / 1+z}^{10 \, \mathrm{MeV} / 1+z} E \, \frac{\mathrm{d}N}{\mathrm{d}E}(E) \, \mathrm{d}E }{
		\int_{20 \, \mathrm{keV}}^{200 \, \mathrm{keV}} E \, \frac{\mathrm{d}N}{\mathrm{d}E}(E) \, \mathrm{d}E }
\end{equation}

where $F_\mathrm{UL}$ is the flux upper-limit calculated as described above, $D_\mathrm{L}$ is the median luminosity distance, and z is the redshift for the event. 
% We report the isotropic equivalent luminosities. 

\section{Results}\label{sec:results}

\subsection{GW event sample}\label{subsec:gwsample}
We used the GW transient sample within the period from the launch of \asat\ (2015 October 6) till the end of the third observing run (O1-3) of LIGO-Virgo GW detectors. This time window leads to the exclusion of GW150914, leaving us with 89 events. All these events meet the thresholds of the event being astrophysical as reported in \citet{gwtc1-2019, gwtc2-2021, gwtc3-2021}. For the analyses in Section~\ref{sec:discussion}, we used the properties, skymaps, and posteriors of the GW events as reported in the respective catalog paper data releases. Apart from these 90 events, more GW events were found by other independent studies \citep{tejaswiNewTrig2020, olsenNewTrig2022}, which are not included in our sample.

Out of these 89 GW events, \asat~was passing through the South Atlantic Anomaly (SAA) for 14 GW events, when CZTI detectors become non-functional. Given the inclination and radius of orbit of \asat, we pass through the SAA region for about 30\% of it. We investigated whether the events for which \asat~was in SAA bring in any bias to the sample. We found that these events are almost evenly distributed across the different mean luminosity distances and hence are not expected to bring any distance bias to our sample. Three GW events --- GW151012, GW200128\_022011, and GW200209\_085452 occurred in a data gap for \czti\, and hence are excluded from our sample.

\begin{sloppypar}
For each of the remaining 72 events, we estimated the instantaneous sky coverage of the skymap, provided in the GWTC data releases, visible to \asat\ given its position and attitude. We sum up the probabilities associated with the pixels that are not occulted by Earth (71\% of the entire sky). These coverage statistics can be found in  Tables~\ref{tab:nstable} and~\ref{tab:bbhtable}. For six events (GW190412, GW190413\_134308, GW190512\_180714,
GW200129\_065458, GW200202\_154313, GW200224\_222234), \asat's coverage was less than 10\%. We do not include these events in further analyses described in  Section~\ref{sec:discussion}. For the special case of GW170817 where the localization of the source was known with arcsecond accuracy due to the discovery of the electromagnetic counterpart, the source was occulted by Earth \citep{kasliwalscience2017} as seen from \asat. However, we still include it in our analysis, using the final GW sky map. 
\end{sloppypar}

For this work, we classify an event as a BNS if both components have masses $m \leq 3~\msun$, NSBH if only one component has mass $\leq 3~\msun$, and BBH otherwise. GW literature often uses a ``Mass Gap'' category for objects in the mass range from $3~\msun$ to $5~\msun$, but we do not consider this as a separate category for our analyses. This classification is on similar lines as in \citet{jointlvkgrb2023}.

\subsection{\czti\ search results}\label{subsec:cztiresults}
Here, we summarise the searches for bursts in \czti\ coincident with the 72 GW events as described above. For each of the 72 events, we search for a coincident GRB-like detection in CZTI data within the search window of 100~s ($\pm$ 50~s) around the merger time. We find no detections in the CZTI data and hence we calculate flux upper limits as described in Section~\ref{sec:dataproc}.

\begin{figure}
	\includegraphics[width=\columnwidth]{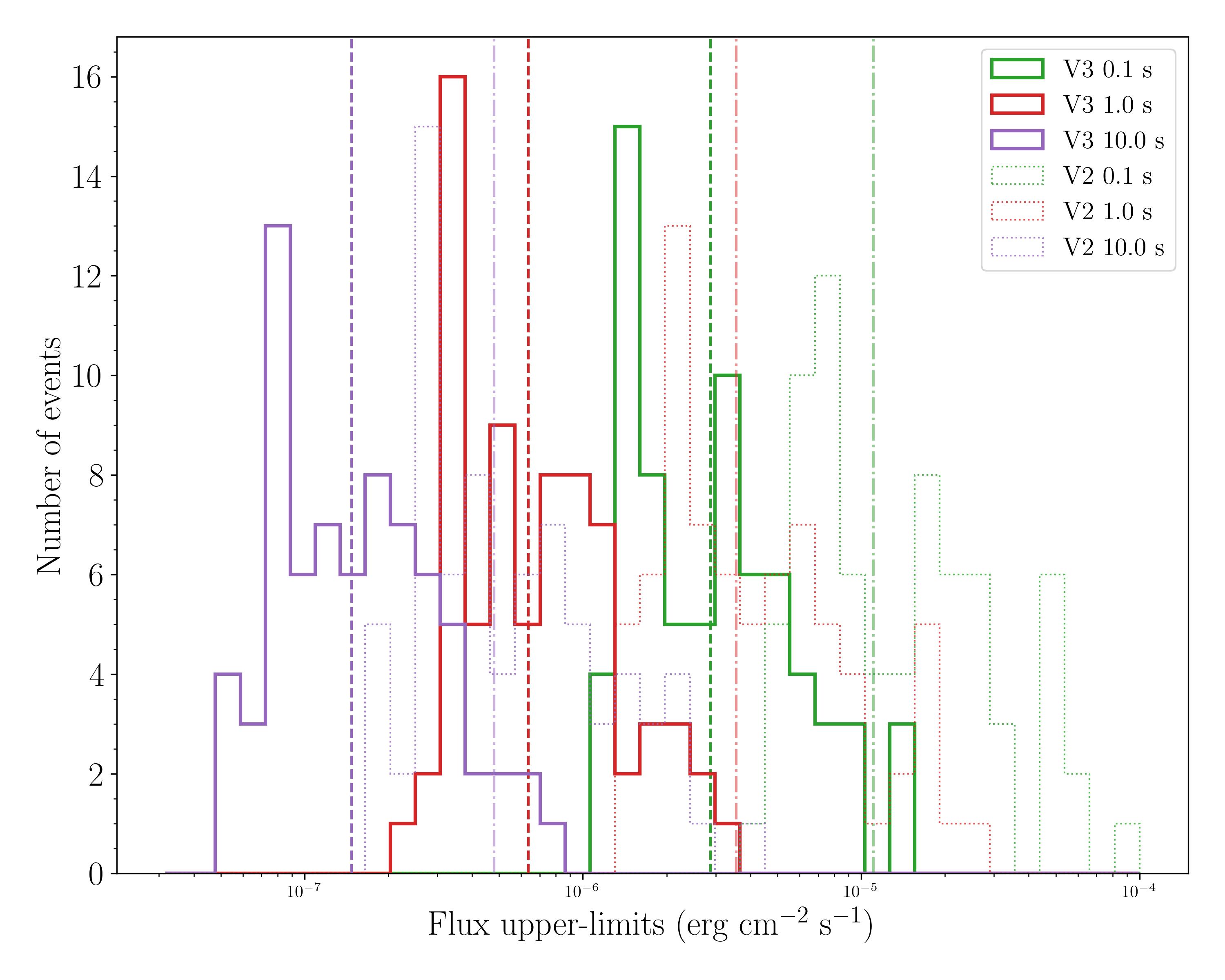}
	\caption{Histogram of the flux upper-limits calculated for all 65 events as described in Section~\ref{subsec:fluxlimits}, for three search time bins: 0.1~s (blue), 1.0~s (orange), and 10.0~s (green). We also show improvement in our estimation of flux upper-limits through the use of CZTI V2 pipeline (dotted) to CZTI V3 pipeline (solid) as described in Section~\ref{sec:dataproc}. The median flux upper-limit for each time bins are also indicated with a dotted line (V2), and a solid line (V3). As you go from lower time bins to higher, the number of samples in the witness window increases, and hence the cutoff count-rate has to increase to achieve the same FAP. As a result, the corresponding flux upper-limits are found to reduce with increasing time bins.}
	\label{fig:flhistogram}
\end{figure}

In Tables~\ref{tab:nstable} and~\ref{tab:bbhtable}, we summarize the 72 GW event sample, their classification as per the mass cuts described in the previous section, skymap coverage for \asat, and the flux upper-limits from our searches. For completeness, these tables also include those events when \asat~was in SAA and the events that do not cross our coverage threshold as described in Section~\ref{subsec:gwsample}. Table~\ref{tab:nstable} contains the flux upper-limits from all events that contain at least one neutron star component ($<~3~\msun$) and Table~\ref{tab:bbhtable} contains all potential BBH merger events, with both component masses $>~3~\msun$. Figure~\ref{fig:flhistogram} shows a histogram of all flux upper-limits from Table~\ref{tab:nstable} and~\ref{tab:bbhtable} at the three different search timescales. We find a median flux upper-limit of 2.87~$\times$~$10^{-6}$~\egs~for 0.1~s, 6.37~$\times$~$10^{-7}$~\egs~for 1.0~s and 1.48~$\times$~$10^{-7}$~\egs~for 10.0~s. 

We report that the flux upper-limits calculated through these methods (described in Section~\ref{sec:dataproc}) improve the limits reported previously through GCNs, e.g. \citet{S191222n, S200219ac}, that were based on the old CZTI V2 pipeline\footnote{\href{http://astrosat-ssc.iucaa.in/cztiData}{http://astrosat-ssc.iucaa.in/cztiData}}. For comparison, Figure~\ref{fig:flhistogram} also shows histograms the flux limits obtained using this older data pipeline, which almost an order of magnitude shallower. We also report that our non-detections are consistent with other analyses \citep{jointgrbgwO1O22020, jointlvkgrb2023, agileGWTC12022, insightGRBGW2021}. We highlight that our different coverage statistics for several of these sources draw different implications on the emission models as discussed in Section~\ref{sec:discussion}.

\section{Discussion}\label{sec:discussion}
We discuss the implications of these non-detections on different emission models for the two classes of events --- events with a neutron star component (Table~\ref{tab:nstable}) and events that are possible BBH mergers (Table~\ref{tab:bbhtable}). We further present the luminosity constraints that can be derived from our non-detections and discuss the prospects from the ongoing fourth observing run (O4) of LVK GW detectors. 

\subsection{Events with a neutron star component}\label{subsec:nscomponent}
\begin{figure}
	\centering
	\includegraphics[width=\linewidth]{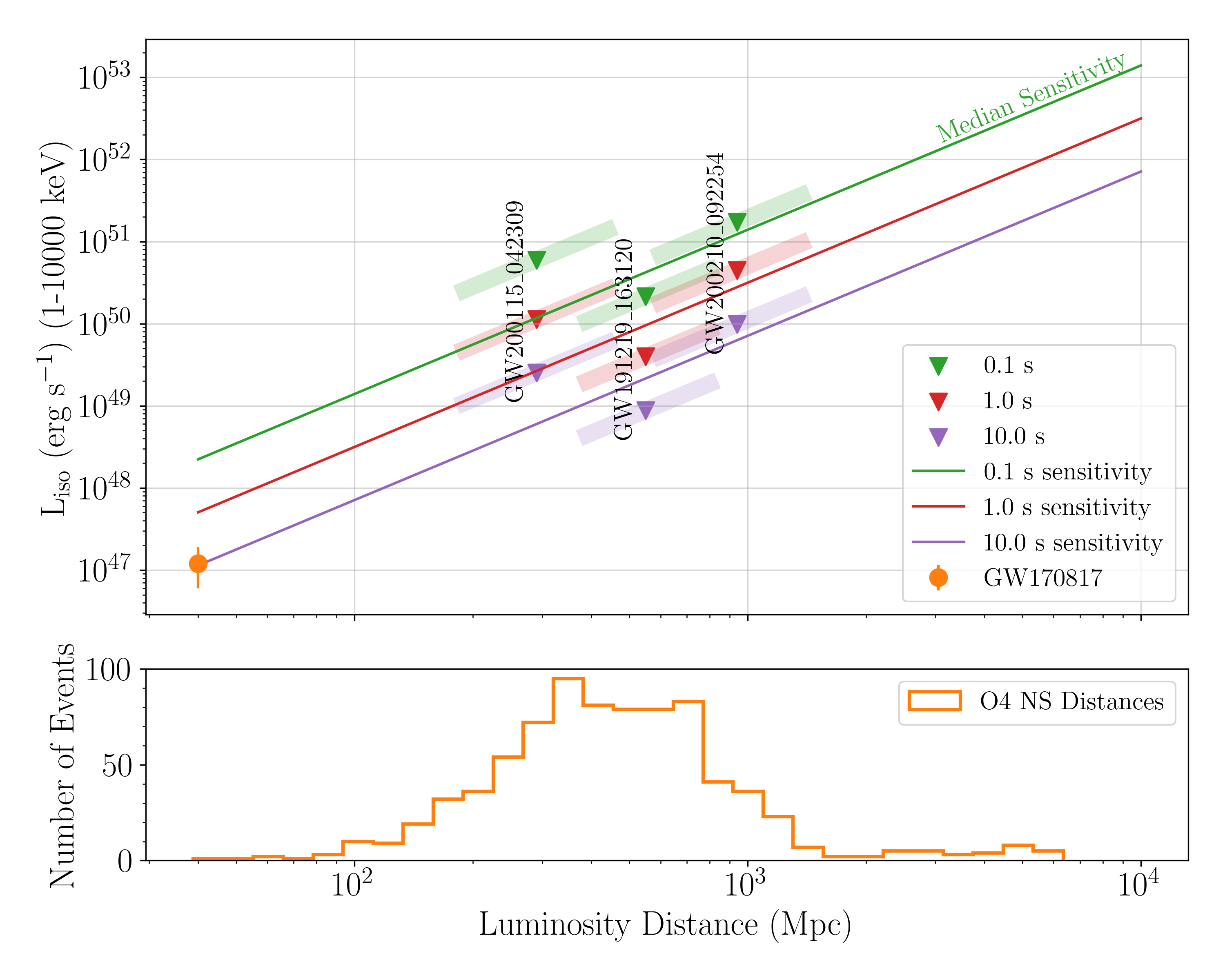}\\
	\caption{Comparison of limits on derived isotropic luminosities for the three events in our GW event sample that contain a neutron star as discussed in Section \ref{subsec:nscomponent}. All values are scaled to the 1~keV -- 10~MeV band. The downward pointing triangles mark the limits on source luminosity at the median distance of these events as given in \citet{gwtc3-2021}, using \czti\ flux upper-limits calculated in Section~\ref{subsec:fluxlimits} for the three time bins: 0.1~s (green), 1.0~s (red), and 10.0~s (purple). The shaded inclined bands show the limits for the 1-$\sigma$ distance ranges for these events. For comparison, the green, red, and purple lines show the median sensitivity of CZTI as a function of distance for searches at these three timescales. Note that due to variations in data quality, the limits for individual events may be deeper or shallower than the median values.
    For comparison, we also show the distance and luminosity of \gfo\ reported by \citet{grbgw2017,bnsmma2017} though it was not detected by CZTI. The orange histogram at the bottom shows the distribution of median distances to the detected NS events from the \citet{weizmann2023} simulations, which gives us an idea of the relevant range of luminosities that CZTI can probe.
	}
	\label{fig:nsplot}
\end{figure}

In our sample of seven GW events (refer Table~\ref{tab:nstable}) that have a potential neutron star component, for three events (GW190917\_114630, GW190814, and GW190425) \asat~was in SAA. For GW170817, the EM counterpart was occulted by Earth for \asat\ --- but we still include the flux upper-limits in Table~\ref{tab:nstable}, with corresponding probability coverage being 0. For the remaining three GW events (GW200210\_092254, GW200115\_042309, and GW191219\_163120), we have reported flux upper-limits. From Figure~\ref{fig:nsplot}, we can see that the flux upper-limits from these three events are roughly consistent with our median sensitivity of CZTI at the three search time bins --- 0.1~s, 1.0~s, and 10.0~s. The median sensitivities, derived from our median flux upper-limits of the complete GW event sample as described in Section~\ref{subsec:gwsample}, also indicate that we are sensitive to fluxes from events like \gfo at 10~s search timescales. We attribute the non-detections of these three events to the fact that they are significantly further away than the measured distance of GW170817 at 40~Mpc \citep{bnsmma2017}, with median distances of about 960~Mpc, 300~Mpc, and 550~Mpc respectively. 

Due to the relatively low number of detected events with a neutron star component, we look ahead toward future observing runs of LVK GW detectors. \citet{petrovdatadriven2022} have shown that even though the sensitivities of the GW detectors will increase, most of the events will be distant, and hence their EM signatures will be faint. They predicted that the combined rate of BNS and NSBH detections would be $\sim 159$ in the 18 months of O4. We caution that this rate would be lower because of Virgo not joining in the first half of O4. In Figure~\ref{fig:o4_bbh_prospects}, we estimate the coverage statistics of \czti\ through the O4 injections set done by \citet{weizmann2023}, which contained 8258 total events (BBH, NSBH, and BNS). We offset the injection times of these events to the years 2018-2020 and used \asat's known orbital positions to estimate whether \asat\ was in SAA during the event, and subsequently calculate the coverage for the skymaps. We note that there are no significant orbital changes in \asat, hence this convenient time shift is an acceptable proxy for the orbital behavior of the satellite in future observing runs. We observe that among 1132 events that have a neutron star component in this injection set, no inferences could be made for 29.5\% events because \asat\ is in SAA, there were data gaps, or the events occulted by Earth for \asat\ (coverage less than 10\%). For the remaining 70.5\% events (corresponding to $\sim 113$ events in the duration of O4), we would have coverage higher than 10\%. Note that the median distance of these ``observable'' events is 435~Mpc, much higher than the \gfo\ distance. In Figure~\ref{fig:nsplot}, we show these events scaled from the luminosity of \gfo\, and we show that they lie just above our 10~s median sensitivity. The histogram (log-scaled) of the median distances of these events is also shown, which shows that the majority of the events would lie much farther away than 100~Mpc. Future missions like {\em Daksha} would be able to detect \gfo\ like events out to a distance of 76~Mpc \citep{dakshatech2022} and hence will be critical in the hunt for EM counterparts from GW. {\em Daksha}, with its all-sky coverage, wide energy band, and an order of magnitude higher sensitivity than CZTI, is expected to detect 12 joint counterparts with LVK every year \citep{dakshascience2022}.

\subsection{Binary Black Hole events}
We have summarized the non-detections of 83 BBH events in Table~\ref{tab:bbhtable}. In this section, we discuss the implications of these non-detections on the luminosity function of these sources, and some models that predict the emission of EM radiation from them.

\subsubsection{Emission models}\label{subsec:bbhmodels}
While BBH mergers are not expected to emit gamma-ray emission due to the lack of matter in the accretion disk, \citet{connaughton2016} reported a 2.9~sigma detection of a gamma-ray signal GRB150914-GBM by \fermi-GBM which was temporally coincident with GW150914. Later, \citet{veres2019} put forth strong constraints on some emission models that provided mechanisms that predict EM emission from BBHs. We refer to the same models for our analysis in this work: a neutrino–antineutrino annihilation powered jet mechanism \citep{yesneutrinoRuffert1998}, the Blandford–Znajek mechanism \citep{yesbzBlandford1977, yesbzKomissarov2010}, and a charged black hole model. These models have been critiqued by others \citep[see for instance][]{noneutrinoLi2016} but are still widely discussed especially in the context of the association of GRB150914-GBM with GW150914  \citep{connaughton2016, jointlvkgrb2023, veres2019}.

\begin{figure}
	\centering
	\includegraphics[width=0.9\columnwidth]{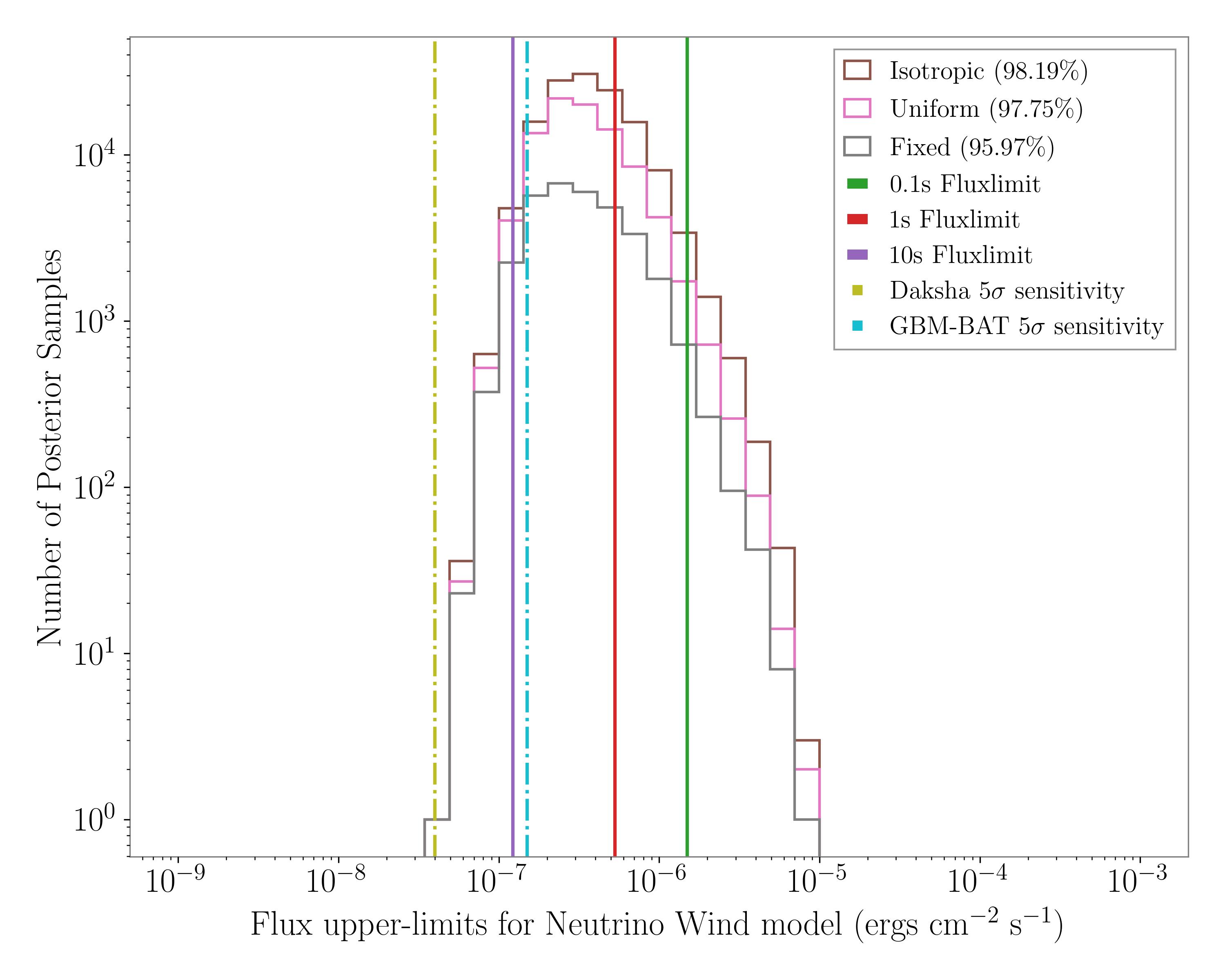}
	\includegraphics[width=0.9\columnwidth]{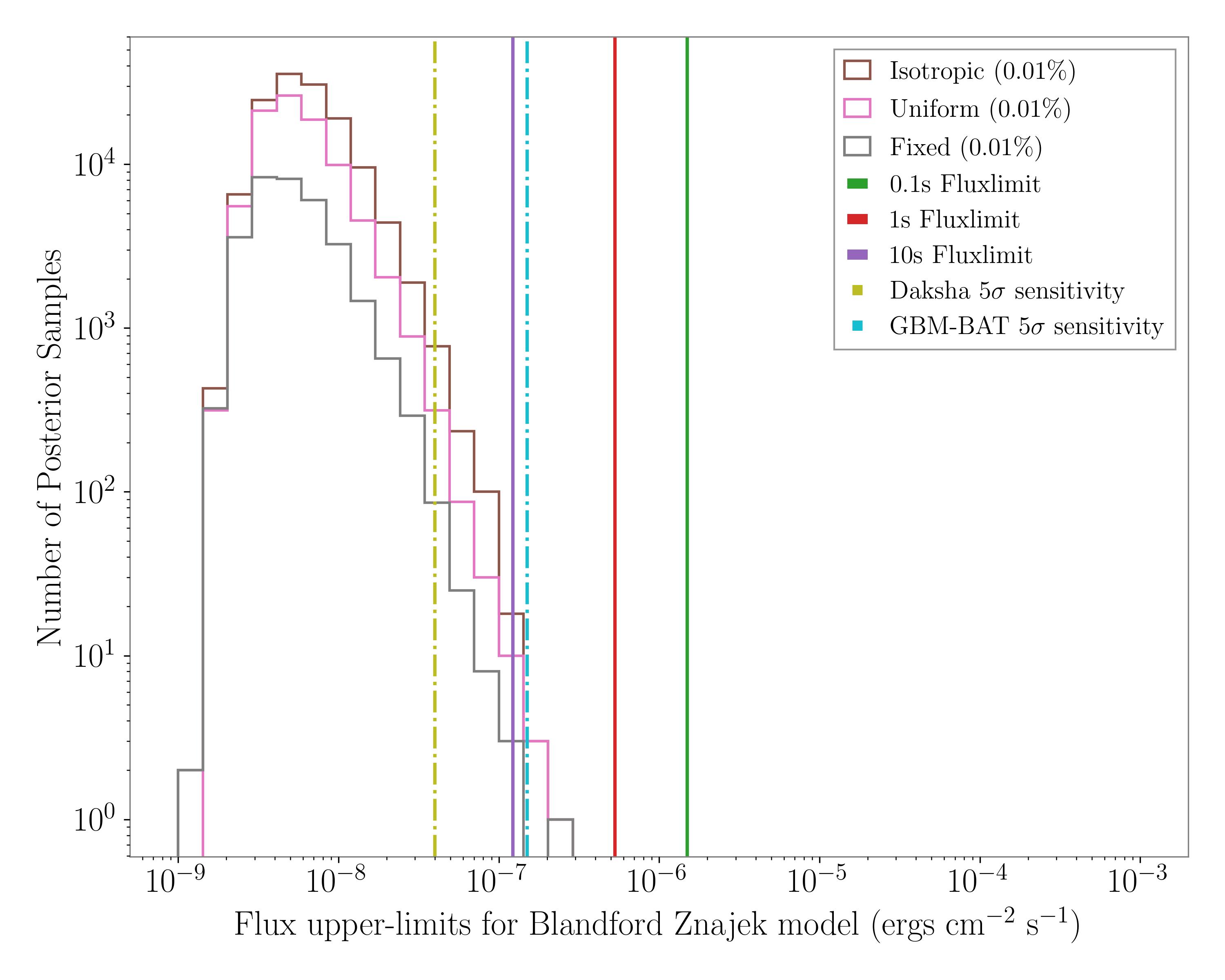}
	\includegraphics[width=0.9\columnwidth]{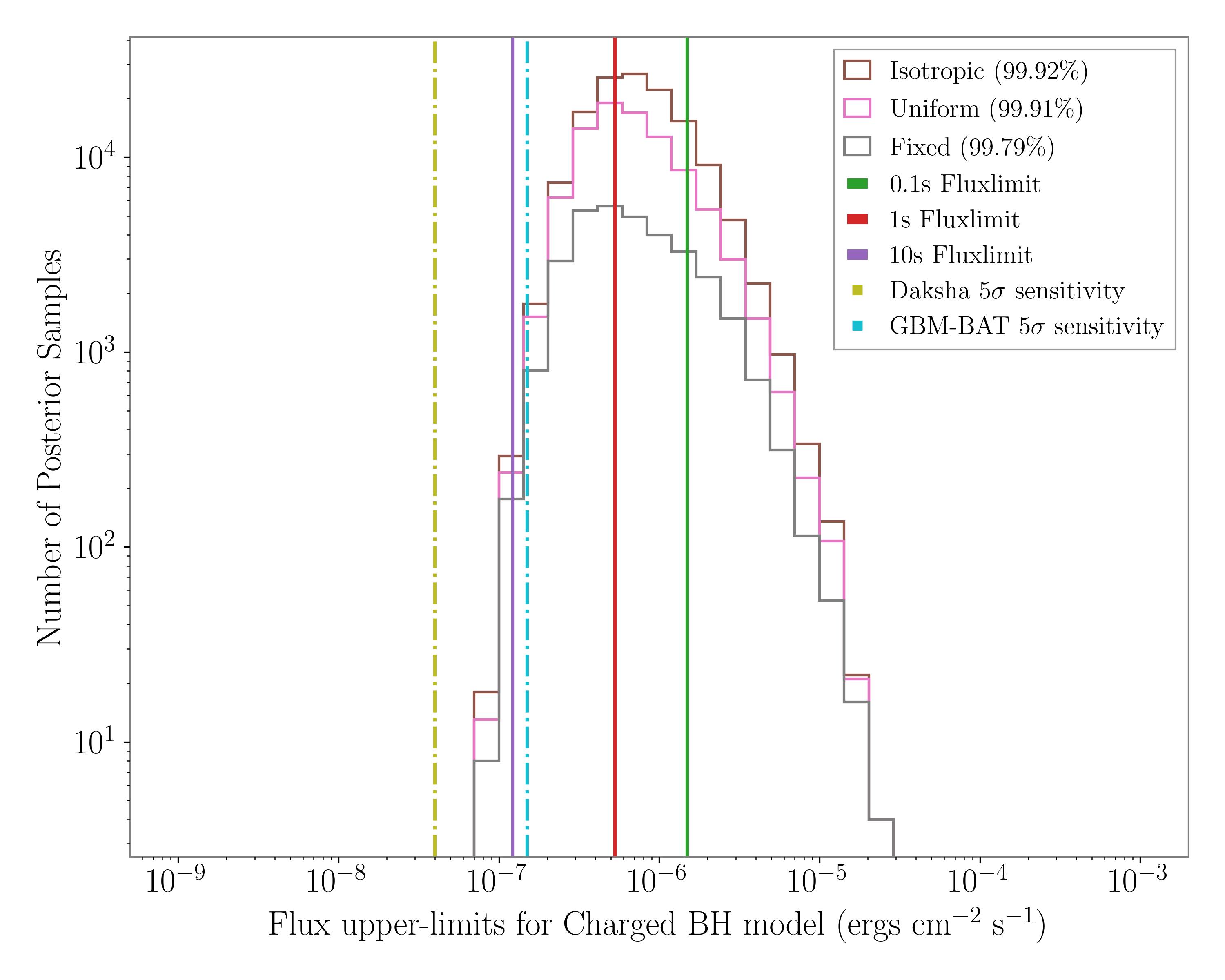}
	\caption{Expected fluxes for GW151226 predicted from the three models --- neutrino–antineutrino annihilation powered jet mechanism, the Blandford Znajek mechanism, and a charged black hole model; that predict EM radiation from BBH mergers, as discussed in Section~\ref{sec:discussion}. The vertical solid lines show our flux upper-limits for this event for the three search time bins, the yellow dashed vertical line shows the 1~s 5-$\sigma$ flux sensitivity for the 20--200~keV band of {\em Daksha} \citep{dakshascience2022}, and the blue dashed vertical line marks the 1~s 5-$\sigma$ sensitivity for the 15--350~keV band search with \swift-BAT and \fermi-GBM \citep{jointlvkgrb2023}. The three histograms within each panel show the three jet models discussed in Section~\ref{subsec:bbhmodels}. For this particular event, our flux upper limits are deep enough to rule out the Neutrino Wind and Charged BH models.}
	\label{fig:gw151226}
\end{figure}

In this section, we explain the methodology to study the implications of our \czti\ flux upper-limits for the BBH non-detections reported in Table~\ref{tab:bbhtable}, which is broadly based upon \citet{veres2019} and \citep{jointgrbgwO1O22020}. Each model used below uses several physical properties of the events. We use one free parameter whose value is derived from GRB150914-GBM observed properties, and the remaining parameters for the model are taken from the physical parameters of the event posteriors. For all these posteriors of an event, we estimate the expected flux for a given model and compare it with the flux upper-limits calculated from our searches. We further consider three cases of GRB jet geometries, through the inclination of the posteriors: an isotropic emitter (no inclination constrain), a uniform emitter (opening angle within the range of 10-40 $\deg$), and a fixed GRB-like jet (opening angle of less than 20 $\deg$). In Figure~\ref{fig:gw151226}, we show how our upper limits compare against the calculated fluxes for all posteriors, grouped by the inclination angles, for a single case of GW151226 as explained below. An important assumption behind the implications stated below is that the source lies within the skymap covered by \asat\ (for instance, the coverage for GW151226 is 95\%). While this assumption needs to be kept in mind while interpreting single-event analyses, in Section~\ref{subsec:gbmconstraints} we undertake a joint evaluation of all non-detections including the coverage of individual events. For comparison with our limits, we highlight the joint--marginal 1~s 5-$\sigma$ sensitivity of \swift-BAT and \fermi-GBM  (15-350~keV) for this event \citep{jointlvkgrb2023} along with the 1~s 5-$\sigma$ sensitivity of the proposed {\em Daksha} mission \citep{dakshascience2022} in the same Figure \ref{fig:gw151226}.

In the neutrino-antineutrino annihilation powered jet model, relativistic jets are launched by the electron-positron pairs formed because of collisions from the neutrino winds from the thick accretion disk around the BBH merger system \citep{yesneutrinoRuffert1998, yesneutrinoZalamea2011, yesneutrinoPage1974}. For this model, the accretion rate is the free parameter which is derived to match the luminosity of the GRB150914-GBM detection. Here we note that there is a typographical error in a term of Equation~4 of \citet{veres2019}, and we have used the correct Equation~15 of \citet{yesneutrinoPage1974}.  We find that for GW151226 our 10~s flux upper-limits are deep enough to rule out more than 98\% of the posteriors of the event (top panel in Figure~\ref{fig:gw151226}).

In the Blandford-Znajek mechanism model, we expect that the rotational energy of the progenitor would be extracted with the help of a magnetic field of the BH \citep{yesbzBlandford1977, yesbzKomissarov2010, yesbzReynolds2006}. Here, the free parameter is the value of the magnetic field, which is scaled to match the GRB150914-GBM luminosity. We find that, unlike the other two models, our flux upper-limits are not deep enough to rule out any particular model, after considering the posterior distributions for each event. Our flux upper-limits are an order of magnitude higher than the median posterior distributions as shown in the middle panel of Figure~\ref{fig:gw151226} for the case of GW151226. Even future higher-sensitivity missions like {\em Daksha} will be able to explore these models only to a limited extent unless the BBH event occurs nearby.

Some models predict that Poynting flux can be extracted to generate gamma-ray emissions from BBHs when one of the black holes is charged \citep{yeschargeZhang2016, yeschargeLui2016}. Here, we scale the charge of the BH to match the luminosity of GRB150914-GBM. For GW151226, our flux upper-limits are deep enough to rule out more than 99\% of the posteriors of this event. We demonstrate an example of the same in the bottom panel of Figure~\ref{fig:gw151226} for the case of GW151226.

\begin{figure}
	\includegraphics[width=\columnwidth]{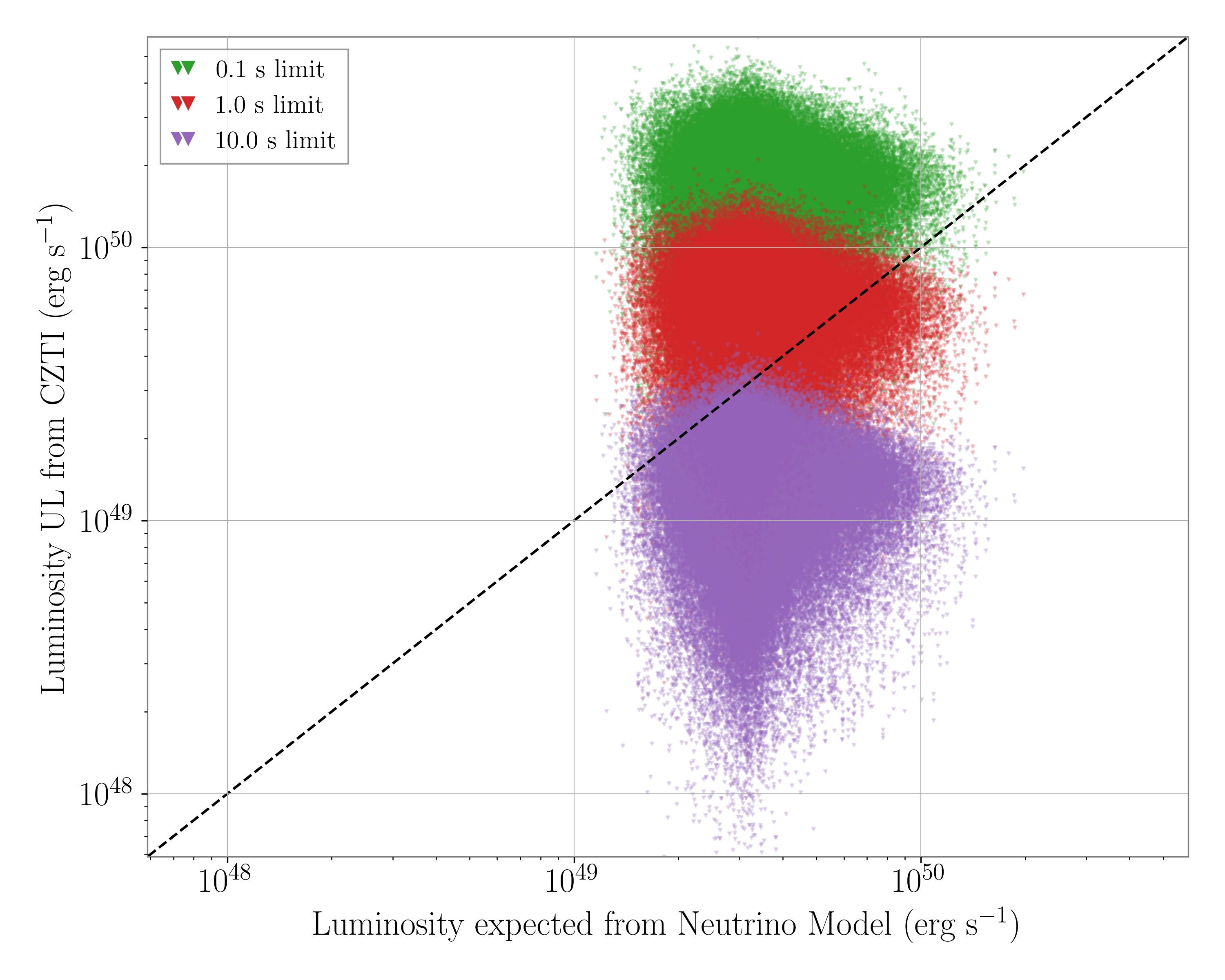}
	\caption{We show the isotropic equivalent luminosity $\mathrm{L_{iso}}$~\eg~(1~keV~-~10~MeV) upper-limits from \czti\ searches as a function of calculated luminosities derived from GRB150914-GBM for the Neutrino Model, for the case of GW151226. As described in Section~\ref{sec:discussion}, we can rule out the luminosities below the diagonal, mainly using our 10~s weighted sensitivity.}
	\label{fig:lumsBBHs}
\end{figure}

We further studied the expected isotropic equivalent luminosities ($\mathrm{L_{iso}}$) (1~keV~-~10~MeV) from the posteriors of each of these events to probe the luminosity parameter space that can be ruled out for each event. We show the same in Figure~\ref{fig:lumsBBHs} for the case of GW151226, for the neutrino model described above, where our flux upper-limits allow us to place constraints on the luminosity space of $10^{49}$ to $2 \times 10^{50}$~\eg, using our weighted 10~s sensitivities for that event.

We note that GW151226, with a median distance of 440~Mpc, is among the relatively nearby BBHs. The majority of the remaining BBHs in our GW sample are much fainter than the current sensitivities of the gamma-ray detectors. While our sensitivity is not high enough to place meaningful constraints on individual event properties, in the next section we explore the implications of a joint analyses of all our non-detections.
% Hence, we emphasize the need to have a deeper flux upper-limit and higher instantaneous coverage of these GW events in order to understand these sources better and further constrain the existing gamma-ray emission models.  
% In Figure~\ref{fig:gw151226}, given the posterior distributions of the triggers seen in the previous observing runs and the lack of EM counterparts, we can see that the sensitivity of a mission like Daksha would play a significant role in the future, and possibly help rule out some of the emission models that predict EM radiation from BBH. 

\subsubsection{GRB150914-GBM Constraints}\label{subsec:gbmconstraints}
\begin{figure}
	\includegraphics[width=\linewidth]{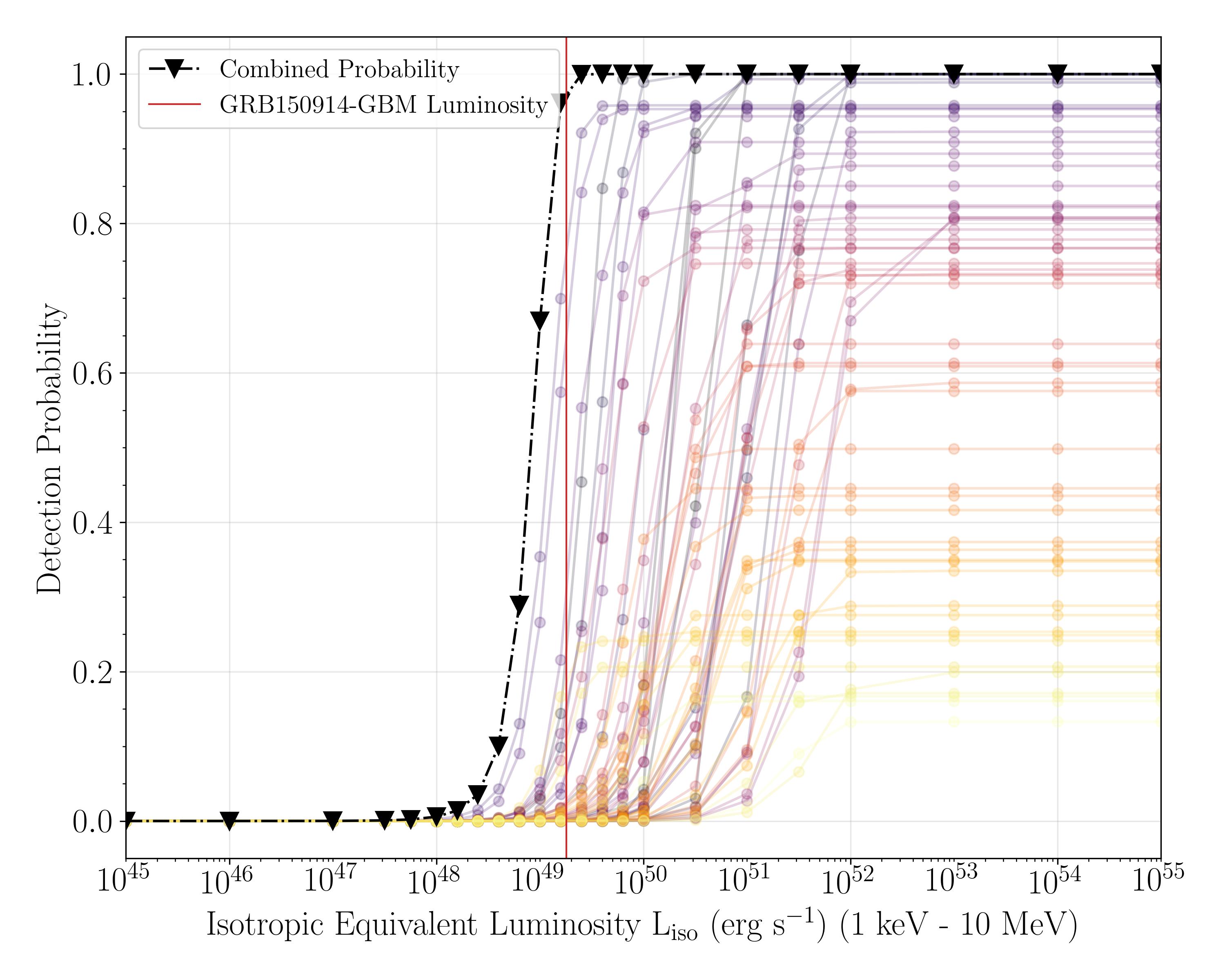}
	\caption{Probability of detection of all 58 BBH events as a function of isotropic equivalent luminosity. The black dashed curve shows the joint detection probability across these 58 BBH events as described in Section~\ref{subsec:gbmconstraints} which shows the detection probability of at least one event out of 58 BBH events. The vertical red line shows the isotropic luminosity of GRB150914-GBM \citep{connaughton2016}, which is the claimed counterpart to GW150914. Through the non-detections of 58 BBHs, all assumed at the luminosity of GRB150914-GBM, we can rule out this luminosity of the claimed counterpart with a confidence of 97\%.}
	\label{fig:nondetectionfull}
\end{figure}

We now estimate the joint probability of \asat~detecting zero out of 58 BBH mergers. For each of these 58 events, we assume a certain source luminosity. Then for each pixel in the Multi-Order-Coverage (MOC) skymap \citep{singerMOC2016}, we calculate the distance to which \czti\ could have detected this source using the flux upper-limit calculated as described in Section \ref{subsec:fluxlimits}. Next, we compute the cumulative probability of detection based on the posterior distributions of each pixel given in the event skymap using the \texttt{conditional\_cdf} function from the \texttt{ligo.skymap.distance} module \citep{singerMOCSup2016}, and multiply this cumulative probability with the GW probability of the pixel from the skymap. We sum these individual probabilities for all pixels in the skymap that are not occulted by Earth for \asat\ to get the combined event detection probability at an assumed luminosity of the source. We repeat this for several luminosities from $10^{45}$ to $10^{55}$~\eg\ as shown in Figure~\ref{fig:nondetectionfull}. We expect that the probability of CZTI detecting an event to saturate to the coverage probability at higher luminosities as the source would be bright enough for CZTI to detect it, and observe the same in Figure~\ref{fig:nondetectionfull}. Note that out of 62 BBH events in our GW sample, we exclude 4 events --- GW190527\_092055, GW190929\_012149, GW200308\_173609, and GW200322\_091133 --- from this analysis since the location parameters of majority pixels in their skymaps (based on the waveform \texttt{IMRPhenomXPHM}) are defined as (\texttt{inf}). The probability in such pixels is extremely small to have a conditional distance distribution represented by \textit{ansatz} \citep{singerMOCSup2016}.

Further, we combine all of these non-detection probabilities to get the probability of detection of at least one of the 58 equal-luminosity BBHs, as a function of luminosity (black--dashed line in Figure~\ref{fig:nondetectionfull}). We note that this assumption of all BBHs having the same luminosity is simplistic, but it is the most basic model-independent case that can be considered. We show that if all 58 BBH mergers were at the luminosity of $3.6 \times 10^{49}$~\eg\ or higher, we can say with a 99.99\% confidence that \czti\ would have detected at least one counterpart among them. Through this, we can also rule out the luminosity of GRB150914-GBM \citep{connaughton2016} with a 97\% confidence, if we were to assume that all 58 BBH events had the same luminosity ($1.8 \times 10^{49}$~\eg) as that of the claimed counterpart of GW150914.

We further observe that the combined non-detection curve in Figure~\ref{fig:nondetectionfull} is dominated by a select few events that are relatively nearby, and well--localised. In our sample, the three closest events are not the most impactful for various reasons. For the closest event GW170608, with a median distance of 320~Mpc, \asat's coverage was relatively low (24\%), and our flux sensitivity at that time was only half the median value. For the event GW191216\_213338, which had a median distance of 340~Mpc, our flux upper-limits were $\sim 4$ times worse than our median flux upper-limits. The third closest event, GW200202\_154313, was fully occulted by the Earth. As a result, the combined non-detection curve closely follows the non-detection plot of GW151226 which, at a distance of 440~Mpc --- much closer than the median value D$_\mathrm{L} \sim 1730$~Mpc for BBH events in our GW sample. This is likely to remain true in the future: the constraints from nearby events will continue to dominate the statistics as compared to the rest of the sample combined. We explore this further in Section~\ref{subsec:o4prospects}.
% Thus, we emphasize the need for well--localised detections of such nearby BBH events, in the future runs of LVK. 
% The addition of more GW detectors like LIGO-India to the IGWN would drastically improve the localisations of events, some by a factor as large as 25 \citep{saleemligoindia2022}, which would be very critical for EM searches in the future.

\subsection{Prospects for O4, and beyond}\label{subsec:o4prospects}
\begin{figure}
	\includegraphics[width=\columnwidth]{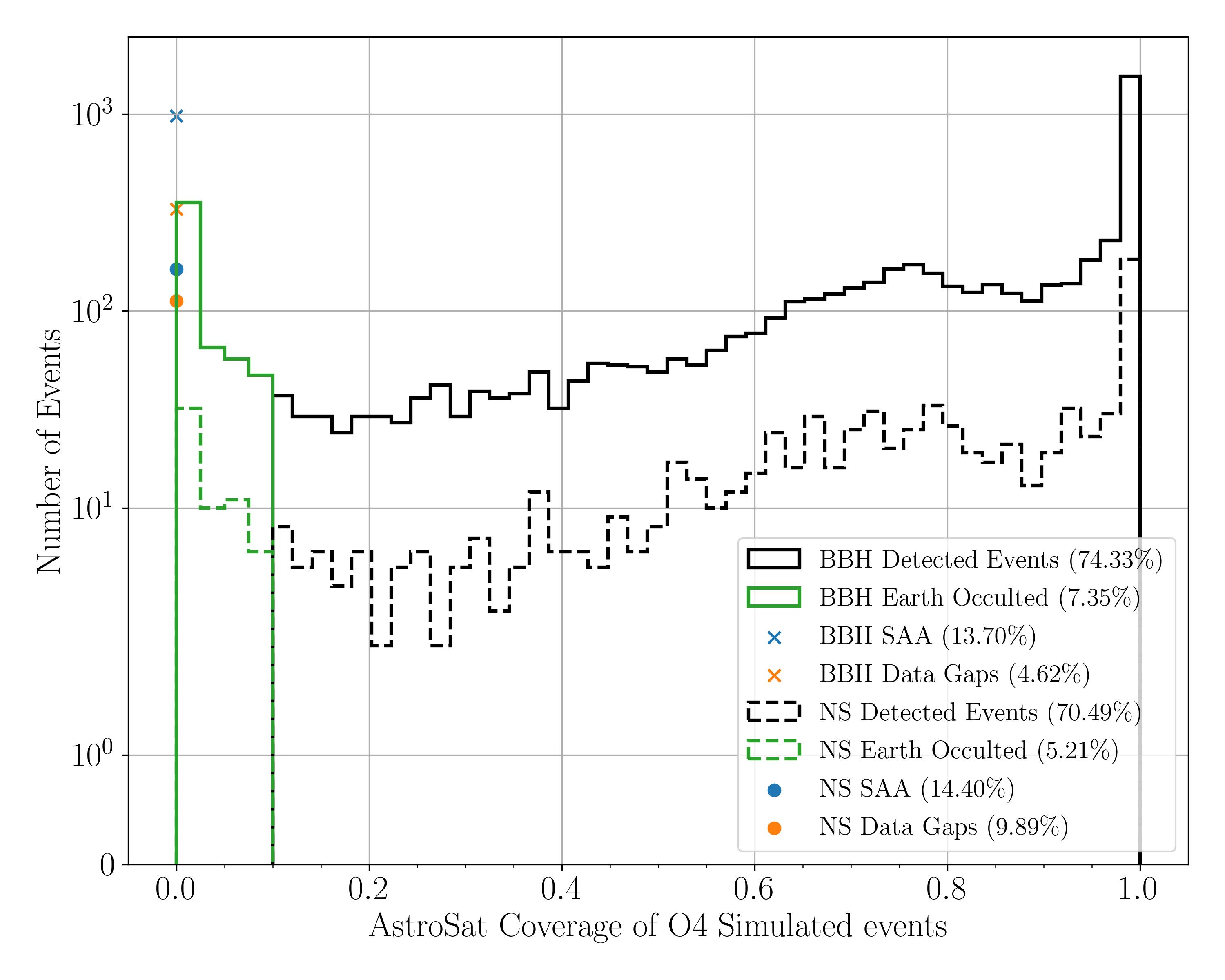}
	\caption{\asat\ coverage statistics for the injection set of the fourth observing run (O4) of LVK GW detectors \citep{weizmann2023} as described in Section~\ref{subsec:o4prospects}. The histogram (solid) shows the recovery of 7126 binary black holes, where we have 74.33\% events are recovered with a coverage higher than 10\% of the skymap (black), while 7.35\% events have a coverage lower than that (green). The other histogram (dashed) shows the recovery of 1132 injections with at least one neutron star component, where we have 70.49\% recovered injections with at least 10\% coverage of the skymap (black), while 5.21\% had a coverage lower than that (green). The blue and orange markers show the events in SAA and Data Gaps respectively for BBH and NS component injections (blue and orange respectively).}
	\label{fig:o4_bbh_prospects}
\end{figure}

With higher expected BBH merger detection rates in the subsequent observing runs of LVK GW detectors, we expect more non-detections given the improved horizon distances of the GW detectors, and lack of comparable increase in the sensitivities of EM detectors \citep{petrovdatadriven2022, weizmann2023}. The number of BBH detections is expected to increase to $\sim$~159 in O4 \citep{petrovdatadriven2022, abbotLRRprospects2020}. We use the same methodology as described in Section \ref{subsec:nscomponent} and use the 7126 O4 BBH injections by \citet{weizmann2023} to simulate \asat's coverage statistics. We again offset the injection times of these events to the years 2018-2020. As shown in Figure~\ref{fig:o4_bbh_prospects}, we have a coverage higher than 10\% for about 74\% or $\sim~118$ events according to the projected rates given in \citet{petrovdatadriven2022}. No inferences could be made for the remaining 25\% due to \asat\ being in SAA, coverage being less than 10\% coverage, or data gaps in CZTI. 

\begin{figure}
	\includegraphics[width=\linewidth]{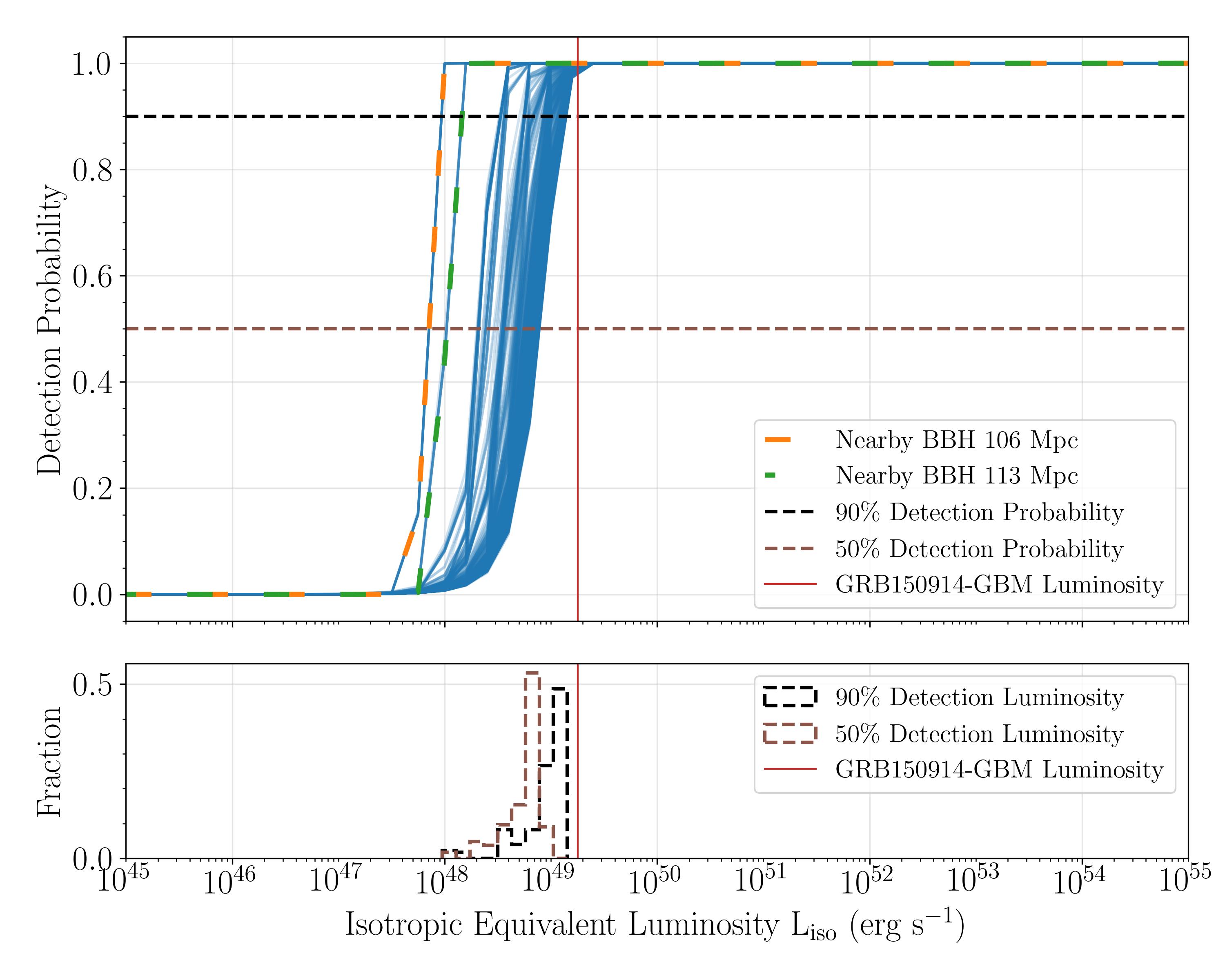}
	\caption{\textit{Top:} Detection probability curves of 500 realisations (blue) of 118 BBH non-detections from the ongoing fourth observing run of LVK GW detectors, combined with the non-detection data from previous runs of LVK GW detectors computed in Section~\ref{subsec:gbmconstraints}. Each of the blue lines shows the joint detection probability of \asat\ detecting at least one out of 118 BBH events predicted in O4, along with the 58 BBH non-detections from previous runs, as described in Section~\ref{subsec:o4prospects}. Bottom: Histogram of the isotropic luminosities we can rule out with 50\% (brown dashed) and 90\% (black dashed) confidence given these 500 realisations. We show that typically these realisations cluster around a $\mathrm{L_{iso}}$ of $10^{49}$~\eg, apart from cases where nearby BBH events (at distances of 106, 113~Mpc) become a part of the realisation, and hence completely dominate the joint-probability curve. In such cases, we would be able to probe even fainter luminosities of $10^{49}$~\eg.}
	\label{fig:o4bbhnondetectionfull}
\end{figure}

We repeat a similar study as shown in Section~\ref{subsec:gbmconstraints} to estimate the luminosities that can be ruled out through these non-detections in O4. For all simulated events, we assume the sensitivity of CZTI as the 10.0~s median flux upper-limit that was obtained in Section~\ref{subsec:cztiresults}. We consider a random subset of 118 BBH non-detections from these 7126 O4 BBH injections and repeat our calculations from this section to create a combined probability curve similar to the dashed black line in Figure~\ref{fig:nondetectionfull}. To avoid any sampling bias, we repeat the procedure a total of 500 times. All these realisations are shown as solid blue lines in the top panel in Figure~\ref{fig:o4bbhnondetectionfull}. The bottom panel of the same figure shows the histogram of isotropic luminosities that we estimate that we can rule out with confidence of 50\% (brown dashed) and 90\% (black dashed). 

Similar to the O1-3 combined non-detection curve (Figure~\ref{fig:nondetectionfull}), the O1-4 combined non-detection curve (Figure~\ref{fig:o4bbhnondetectionfull}) is also heavily dominated by nearby well-localised sources. We draw attention to two such BBH mergers -- at a distance of 106~Mpc and 113~Mpc. The individual non-detection curves for these are shown by dashed orange and dashed green lines respectively. We find that multiple realisations overlap with these lines. The explanation is similar to what we observed with GW151226 in Section~\ref{subsec:gbmconstraints}: whenever our 118-source subset had any of these events present, their luminosity constraints were far stronger than the rest of the sample, causing the combined curve to nearly overlap with the individual source curve. In these cases, we can rule out a BBH merger luminosity of $10^{48}$~\eg\ with 90\% confidence.

% Majority of our realisations cluster around the $10^{49}$~\eg\ since most of the new BBH detections in O4 are expected to be distant. In the few cases where the realisations contain either of these two relatively nearby BBH mergers, we are able to rule out $10^{48}$~\eg\ with a confidence of 90\%, as shown in the bottom panel of Figure~\ref{fig:o4bbhnondetectionfull}. 

%\ifh{show an extrapolated plot of joint non-detection for X BBHs from O4.}

\section{Summary}
Over the three past observing runs, LVK GW detectors have detected 90 CBC GW triggers, out of which 83 were potential BBHs, and the remaining 7 had a neutron star component (either a BNS or an NSBH). We undertook a systematic search for hard X-ray bursts in \czti\ data coincident with these triggers. For the events during which CZTI detectors were functioning, we found no coincident detections. We calculated GW-skymap probability weighted X-ray flux upper-limits in the operating range of CZTI (20-200~keV) for all these events.

For the three mergers in our sample that potentially contained a neutron star component, the luminosity distance of the mergers was too large to have produced any detectable flux given the median sensitivity of \czti. For these three events, the partial Earth occultation of the coverage skymap also meant that we could have missed the real event. 

The majority of our GW sample comprised of BBHs, accounting for 58 events. We use their non-detections to constrain the emission models. We saw that for a relatively nearby BBH event GW151226, at a distance of 440~Mpc, we were able to rule out the neutrino–antineutrino annihilation powered jet mechanism (95\%) and the emission due to a charged BH model (99\%), assuming that the source was not occulted by earth for \asat. We further estimated the detection probability of 58 of these events that had good distance posteriors, and a combined probability of detecting at least one out of these 58 BBHs. We show that we can rule out an isotropic luminosity of $3.6 \times 10^{49}$~\eg\ with a confidence of 99.99\%. Through the same estimate, with a confidence of 97\%, we can rule out the possibility that all BBH mergers had the luminosity of GRB150914-GBM, the putative counterpart to GW150914. 

We expect a higher number of BBH mergers in the ongoing fourth observing run (O4) of LVK GW detectors. By calculating the combined probability of non-detections of all BBH events in O4, we estimate to rule out a luminosity of $10^{49}$~\eg\ with a 90\% confidence. If LVK GW detectors can detect a few relatively nearby BBH mergers (around 100~Mpc), then we would be able to probe a luminosity as low as $10^{48}$~\eg\ as well. Thus, we highlight the importance of the detection of nearby BBH mergers. 

To improve our estimates of \asat-CZTI's flux upper-limits, which currently are quite conservative, we are currently developing a search pipeline that would further improve \czti's flux upper-limit estimates. \czti\ continues to carry out automated searches for EM counterparts for GW in the currently ongoing observing run O4 which are reported regularly in the GCN circulars and on our webpage\footnote{\url{http://astrosat.iucaa.in/czti/?q=emgw}}. Our results from this observing run will be summarised after the end of O4.

To better constrain the BBH emission models and to probe a deeper isotropic luminosity, we emphasise the need for a more sensitive X-ray all-sky monitor suitable for GW counterpart searches, to make up for the increased sensitivity of the GW detectors. Future missions like {\em Daksha} would be ideal to further constrain the BBH emission models discussed in this work, potentially rule them out consistently, and probe even fainter luminosities. 

\section{Software and third party data repository citations} \label{sec:softwarecite}
% Astropy, ligo-skymap, gracedb, scipy.
This work utilised various software including Python \citep{python3}, AstroPy \citep{astropy}, NumPy \citep{numpy}, and Matplotlib \citep{matplotlib}. We used the public data release of the three GWTCs for skymaps, posterior samples, and properties of all individual events, hosted on Zenodo \citep{zenodo}. This research has made use of NASA’s Astrophysics Data System.

\section{Acknowledgments}
We thank Matthew Graham, Sanjit Mitra, A. R. Rao, and Gulab Dewangan for fruitful discussions regarding the analyses. We also thank Vedant Shenoy for his contributions to the initial setting up of the codes that were then further developed by the current team. Lastly, we thank the CZTI Interface for Fast Transients (CIFT) Team at IIT Bombay, and the Payloads Operations Center (POC) team at IUCAA, Pune, for their valuable contributions to the regular transient searches in the \asat-CZTI data. We also thank Vishwajeet Swain, Utkarsh Pathak, Anirudh Salgundi, and Anuraag Arya for their valuable comments on the paper text. 

CZT--Imager is built by a consortium of Institutes across India. The Tata Institute of Fundamental Research, Mumbai, led the effort with instrument design and development. Vikram Sarabhai Space Centre (VSSC) at Thiruvananthapuram provided the electronic design, assembly, and testing. ISRO Satellite Centre (ISAC) at Bengaluru provided the mechanical design, quality consultation, and project management. The Inter-University Centre for Astronomy and Astrophysics (IUCAA) at Pune did the Coded Mask design, instrument calibration, and the Payload Operation Centre. Space Application Centre (SAC) at Ahmedabad provided the analysis software. Physical Research Laboratory (PRL) Ahmedabad, provided the polarisation detection algorithm and ground calibration. A vast number of industries participated in the fabrication and the University sector pitched in by participating in the test and evaluation of the payload.

The Indian Space Research Organisation (ISRO) funded, managed, and facilitated the project.

%%%%%%%%%%%%%%%%%%%% REFERENCES %%%%%%%%%%%%%%%%%%

\bibliography{ref}{}
\bibliographystyle{apj}

%%%%%%%%%%%%%%%%%%%% Tables %%%%%%%%%%%%%%%%%%

\clearpage

\startlongtable
\begin{deluxetable}{cccccchccccc}
\tabletypesize{\small}
\tablecaption{\czti\ results for GW events, with at least one neutron star component (M$_2 < 3 \msun$), from the first three observing runs of the LIGO-Virgo GW detectors. The column labeled D$_\mathrm{l}$, z, M$_1$, and M$_2$ correspond to the luminosity distance, redshift, mass of the more massive component of the binary and the mass of the less massive component in the binary, respectively. The Merger column corresponds to whether both the components of the binary are neutron stars (BNS) or just a neutron star (NSBH), based on the 3~\msun\ mass threshold. The columns --- F$_\mathrm{UL 0.1}$, F$_\mathrm{UL 1.0}$, F$_\mathrm{UL 10.0}$ are the flux upper-limits for the three search time bins 0.1~s, 1.0~s, and 10.0~s respectively. The column labeled SAA shows the status of \asat\ being in the South Atlantic Anomaly (SAA), and the last column labeled Coverage corresponds to the probability covered by \asat\ for each GW event. 
	\label{tab:nstable}}
\tablehead{
%\colhead{Name}  &  \colhead{D$_{\mathrm{l}}$}  &  \colhead{z}  &  \colhead{M$_{1}$}  &  \colhead{M$_{2}$}  &  \colhead{Merger}  &  \nocolhead{p$_{\mathrm{astro}}$}  &  \colhead{F$_\mathrm{UL 0.1}$}  &  \colhead{F$_\mathrm{UL 1.0}$}  &  \colhead{F$_\mathrm{UL 10.0}$}  &  \colhead{SAA}  &  \colhead{Coverage} \\
%	& (Mpc)  &  & (\msun)  &  (\msun)  &  &  & \multicolumn{3}{c}{ ($10^{-7}~\mathrm{erg}\,\mathrm{cm}^{-2}\,\mathrm{s}^{-1}$)}  &  & 
Name  &  D$_\mathrm{L}$  &  z  &  M$_1$  &  M$_2$  &  Merger  &  p$_\mathrm{astro}$  &  \multicolumn{3}{c}{F$_\mathrm{UL}$ ($10^{-7}~\mathrm{erg}\,\mathrm{cm}^{-2}\,\mathrm{s}^{-1}$)} &  SAA  &  Coverage \\
\cline{7-10}
	& (Mpc)  &  & (\msun)  &  (\msun)  &  &  & 0.1~s & 1.0~s & 10.0s  &  & 
}
\decimals
\startdata
GW170817 & 40$^{+7}_{-15}$  & 0.01$^{+0.00}_{-0.00}$ & 1.46 & 1.27 & BNS & 1.00 & 62.04 & 11.97 & 3.77 & No & 0.00  \\
GW190425 & 150$^{+80}_{-60}$ & 0.03$^{+0.02}_{-0.01}$ & 2.1  & 1.3  & BNS & 0.78 & \nodata & \nodata & \nodata & Yes & \nodata \\
GW190814 & 230$^{+40}_{-50}$ & 0.05$^{+0.01}_{-0.01}$ & 23.3 & 2.6  & NSBH & 1.00 & \nodata & \nodata & \nodata & Yes & \nodata \\
GW190917\_114630 & 720$^{+300}_{-310}$ & 0.15$^{+0.05}_{-0.06}$ & 9.7  & 2.1  & NSBH & 0.77 & \nodata & \nodata & \nodata & Yes & \nodata \\
GW191219\_163120 & 550$^{+250}_{-160}$ & 0.11$^{+0.05}_{-0.03}$ & 31.1 & 1.17 & NSBH & 0.82 & 14.87 & 2.74 & 0.61 & No & 0.60  \\
GW200115\_042309 & 290$^{+150}_{-100}$ & 0.06$^{+0.03}_{-0.02}$ & 5.9  & 1.44 & NSBH & 0.99 & 148.45 & 28.01 & 6.29 & No & 0.18  \\
GW200210\_092254 & 940$^{+430}_{-340}$ & 0.19$^{+0.08}_{-0.06}$ & 24.1 & 2.83 & NSBH & 0.42 & 40.68 & 10.61 & 2.33 & No & 0.38  \\
\enddata
\end{deluxetable}

\startlongtable
\begin{deluxetable*}{cccccchccccc}
\tabletypesize{\small}
\tablecaption{\czti\ results for GW events, that are binary black holes (BBHs) (M$_2 > 3 \mathrm{M}_{\odot}$), from the first three observing runs of the LIGO-Virgo GW detectors. The column labeled D$_\mathrm{l}$, z, M$_1$, and M$_2$ correspond to the luminosity distance, redshift, mass of the more massive component of the binary and the mass of the less massive component in the binary, respectively. The Merger column corresponds to whether both the components of the binary are neutron stars (BNS) or just a neutron star (NSBH), based on the 3 \msun mass threshold. The columns --- FL$_{0.1}$, FL$_{1.0}$, FL$_{10.0}$ are the flux upper-limits for the three search time bins 0.1~s, 1.0~s, and 10.0~s respectively. The column labeled SAA shows the status of \asat\ being in the South Atlantic Anomaly (SAA), and the last column labeled Coverage corresponds to the probability covered by \asat\ for each GW event. 
	\label{tab:bbhtable}}
\tablehead{
Name  &  D$_\mathrm{L}$  &  z  &  M$_1$  &  M$_2$  &  Merger  &  p$_\mathrm{astro}$  &  \multicolumn{3}{c}{F$_\mathrm{UL}$ ($10^{-7}~\mathrm{erg}\,\mathrm{cm}^{-2}\,\mathrm{s}^{-1}$)} &  SAA  &  Coverage \\
\cline{7-10}
	& (Mpc)  &  & (\msun)  &  (\msun)  &  &  & 0.1~s & 1.0~s & 10.0s  &  & 
}
\startdata
GW150914 & 440$^{+150}_{-170}$ & 0.09$^{+0.03}_{-0.03}$ & 35.6 & 30.6 & BBH & 1.00 & \nodata & \nodata & \nodata & No & \nodata \\
GW151012 & 1080$^{+550}_{-490}$ & 0.21$^{+0.09}_{-0.09}$ & 23.2 & 13.6 & BBH & 1.00 & \nodata & \nodata & \nodata & No & \nodata \\
GW151226 & 450$^{+180}_{-190}$ & 0.09$^{+0.04}_{-0.04}$ & 13.7 & 7.7 & BBH & 1.00 & 14.97 & 5.31 & 1.23 & No & 0.96 \\
GW170104 & 990$^{+440}_{-430}$ & 0.20$^{+0.08}_{-0.08}$ & 30.8 & 20.0 & BBH & 1.00 & 19.07 & 3.54 & 0.79 & No & 1.00 \\
GW170608 & 320$^{+120}_{-110}$ & 0.07$^{+0.02}_{-0.02}$ & 11.0 & 7.6 & BBH & 1.00 & 37.34 & 9.50 & 2.07 & No & 0.24 \\
GW170729 & 2840$^{+1400}_{-1360}$ & 0.49$^{+0.19}_{-0.21}$ & 50.2 & 34.0 & BBH & 0.94 & 14.16 & 3.87 & 1.33 & No & 0.72 \\
GW170809 & 1030$^{+320}_{-390}$ & 0.20$^{+0.05}_{-0.07}$ & 35.0 & 23.8 & BBH & 1.00 & \nodata & \nodata & \nodata & Yes & \nodata \\
GW170814 & 600$^{+150}_{-220}$ & 0.12$^{+0.03}_{-0.04}$ & 30.6 & 25.2 & BBH & 1.00 & 20.83 & 5.32 & 1.16 & No & 0.21 \\
GW170818 & 1060$^{+420}_{-380}$ & 0.21$^{+0.07}_{-0.07}$ & 35.4 & 26.7 & BBH & 1.00 & 15.53 & 3.86 & 0.83 & No & 0.95 \\
GW170823 & 1940$^{+970}_{-900}$ & 0.35$^{+0.15}_{-0.15}$ & 39.5 & 29.0 & BBH & 1.00 & 16.96 & 3.33 & 0.73 & No & 0.50 \\
GW190403\_051519 & 8280$^{+6720}_{-4290}$ & 1.18$^{+0.73}_{-0.53}$ & 85.0 & 20.0 & BBH & 0.61 & 14.62 & 6.60 & 0.95 & No & 0.81 \\
GW190408\_181802 & 1540$^{+440}_{-620}$ & 0.29$^{+0.07}_{-0.11}$ & 24.8 & 18.5 & BBH & 1.00 & 136.08 & 34.89 & 7.71 & No & 1.00 \\
GW190412 & 720$^{+240}_{-220}$ & 0.15$^{+0.04}_{-0.04}$ & 27.7 & 9.0 & BBH & 1.00 & 13.66 & 2.58 & 0.57 & No & 0.01 \\
GW190413\_052954 & 3320$^{+1910}_{-1400}$ & 0.56$^{+0.25}_{-0.21}$ & 33.7 & 24.2 & BBH & 0.93 & 39.57 & 10.41 & 2.34 & No & 0.58 \\
GW190413\_134308 & 3800$^{+2480}_{-1830}$ & 0.62$^{+0.32}_{-0.26}$ & 51.3 & 30.4 & BBH & 0.99 & 45.57 & 11.91 & 2.68 & No & 0.09 \\
GW190421\_213856 & 2590$^{+1490}_{-1240}$ & 0.45$^{+0.21}_{-0.19}$ & 42.0 & 32.0 & BBH & 1.00 & 14.52 & 3.70 & 0.79 & No & 0.44 \\
GW190426\_190642 & 4580$^{+3400}_{-2280}$ & 0.73$^{+0.41}_{-0.32}$ & 105.5 & 76.0 & BBH & 0.75 & \nodata & \nodata & \nodata & Yes & \nodata \\
GW190503\_185404 & 1520$^{+630}_{-600}$ & 0.29$^{+0.10}_{-0.10}$ & 41.3 & 28.3 & BBH & 1.00 & 19.82 & 3.76 & 0.83 & No & 0.17 \\
GW190512\_180714 & 1460$^{+510}_{-590}$ & 0.28$^{+0.08}_{-0.10}$ & 23.2 & 12.5 & BBH & 1.00 & 19.79 & 3.73 & 0.82 & No & 0.01 \\
GW190513\_205428 & 2210$^{+990}_{-810}$ & 0.40$^{+0.14}_{-0.13}$ & 36.0 & 18.3 & BBH & 1.00 & 14.57 & 3.66 & 0.61 & No & 0.16 \\
GW190514\_065416 & 3890$^{+2610}_{-2070}$ & 0.64$^{+0.33}_{-0.30}$ & 40.9 & 28.4 & BBH & 0.76 & 15.84 & 5.46 & 0.86 & No & 0.81 \\
GW190517\_055101 & 1790$^{+1750}_{-880}$ & 0.33$^{+0.26}_{-0.15}$ & 39.2 & 24.0 & BBH & 1.00 & 12.77 & 2.43 & 0.54 & No & 0.99 \\
GW190519\_153544 & 2600$^{+1720}_{-960}$ & 0.45$^{+0.24}_{-0.15}$ & 65.1 & 40.8 & BBH & 1.00 & 12.50 & 3.28 & 0.54 & No & 0.61 \\
GW190521 & 3310$^{+2790}_{-1800}$ & 0.56$^{+0.36}_{-0.27}$ & 98.4 & 57.2 & BBH & 1.00 & 84.64 & 21.15 & 3.49 & No & 0.73 \\
GW190521\_074359 & 1080$^{+580}_{-530}$ & 0.21$^{+0.10}_{-0.10}$ & 43.4 & 33.4 & BBH & 1.00 & \nodata & \nodata & \nodata & Yes & \nodata \\
GW190527\_092055 & 2520$^{+2080}_{-1230}$ & 0.44$^{+0.29}_{-0.19}$ & 35.6 & 22.2 & BBH & 0.85 & 12.02 & 3.76 & 1.09 & No & 0.65 \\
GW190602\_175927 & 2840$^{+1930}_{-1280}$ & 0.49$^{+0.26}_{-0.20}$ & 71.8 & 44.8 & BBH & 1.00 & 23.93 & 6.14 & 1.37 & No & 0.36 \\
GW190620\_030421 & 2910$^{+1710}_{-1320}$ & 0.50$^{+0.23}_{-0.20}$ & 58.0 & 35.0 & BBH & 0.99 & 19.04 & 7.97 & 2.17 & No & 0.88 \\
GW190630\_185205 & 870$^{+530}_{-360}$ & 0.18$^{+0.09}_{-0.07}$ & 35.1 & 24.0 & BBH & 1.00 & 16.54 & 4.63 & 1.00 & No & 0.77 \\
GW190701\_203306 & 2090$^{+770}_{-740}$ & 0.38$^{+0.11}_{-0.12}$ & 54.1 & 40.5 & BBH & 1.00 & 30.02 & 5.68 & 1.71 & No & 1.00 \\
GW190706\_222641 & 3630$^{+2600}_{-2000}$ & 0.60$^{+0.33}_{-0.29}$ & 74.0 & 39.4 & BBH & 1.00 & 45.90 & 11.85 & 2.57 & No & 0.13 \\
GW190707\_093326 & 850$^{+340}_{-400}$ & 0.17$^{+0.06}_{-0.08}$ & 12.1 & 7.9 & BBH & 1.00 & 28.30 & 5.46 & 1.23 & No & 0.82 \\
GW190708\_232457 & 930$^{+310}_{-390}$ & 0.19$^{+0.06}_{-0.07}$ & 19.8 & 11.6 & BBH & 1.00 & 32.83 & 8.49 & 1.87 & No & 0.45 \\
GW190719\_215514 & 3730$^{+3120}_{-2070}$ & 0.61$^{+0.39}_{-0.30}$ & 36.6 & 19.9 & BBH & 0.92 & 142.12 & 27.03 & 6.24 & No & 0.81 \\
GW190720\_000836 & 770$^{+650}_{-260}$ & 0.16$^{+0.11}_{-0.05}$ & 14.2 & 7.5 & BBH & 1.00 & 13.50 & 3.43 & 0.79 & No & 0.94 \\
GW190725\_174728 & 1030$^{+520}_{-430}$ & 0.20$^{+0.09}_{-0.08}$ & 11.8 & 6.3 & BBH & 0.96 & 14.47 & 3.78 & 0.86 & No & 0.91 \\
GW190727\_060333 & 3070$^{+1300}_{-1230}$ & 0.52$^{+0.18}_{-0.18}$ & 38.9 & 30.2 & BBH & 1.00 & 83.74 & 21.02 & 4.68 & No & 0.92 \\
GW190728\_064510 & 880$^{+260}_{-380}$ & 0.18$^{+0.05}_{-0.07}$ & 12.5 & 8.0 & BBH & 1.00 & 48.74 & 12.26 & 2.76 & No & 0.25 \\
GW190731\_140936 & 3330$^{+2350}_{-1770}$ & 0.56$^{+0.31}_{-0.26}$ & 41.8 & 29.0 & BBH & 0.83 & \nodata & \nodata & \nodata & Yes & \nodata \\
GW190803\_022701 & 3190$^{+1630}_{-1470}$ & 0.54$^{+0.22}_{-0.22}$ & 37.7 & 27.6 & BBH & 0.97 & 69.74 & 18.13 & 3.96 & No & 1.00 \\
GW190805\_211137 & 6130$^{+3720}_{-3080}$ & 0.92$^{+0.43}_{-0.40}$ & 46.2 & 30.6 & BBH & 0.95 & 34.72 & 8.79 & 1.96 & No & 0.20 \\
GW190828\_063405 & 2070$^{+650}_{-920}$ & 0.38$^{+0.10}_{-0.15}$ & 31.9 & 25.8 & BBH & 1.00 & 87.81 & 16.82 & 3.82 & No & 0.73 \\
GW190828\_065509 & 1540$^{+690}_{-650}$ & 0.29$^{+0.11}_{-0.11}$ & 23.7 & 10.4 & BBH & 1.00 & 58.75 & 11.25 & 2.56 & No & 0.35 \\
GW190910\_112807 & 1520$^{+1090}_{-630}$ & 0.29$^{+0.17}_{-0.11}$ & 43.8 & 34.2 & BBH & 1.00 & 26.98 & 6.95 & 1.53 & No & 0.64 \\
GW190915\_235702 & 1750$^{+710}_{-650}$ & 0.32$^{+0.11}_{-0.11}$ & 32.6 & 24.5 & BBH & 1.00 & 31.97 & 6.07 & 1.31 & No & 1.00 \\
GW190916\_200658 & 4940$^{+3710}_{-2380}$ & 0.77$^{+0.45}_{-0.32}$ & 43.8 & 23.3 & BBH & 0.66 & \nodata & \nodata & \nodata & Yes & \nodata \\
GW190924\_021846 & 550$^{+220}_{-220}$ & 0.11$^{+0.04}_{-0.04}$ & 8.8 & 5.1 & BBH & 1.00 & \nodata & \nodata & \nodata & Yes & \nodata \\
GW190925\_232845 & 930$^{+460}_{-350}$ & 0.19$^{+0.08}_{-0.07}$ & 20.8 & 15.5 & BBH & 0.99 & \nodata & \nodata & \nodata & Yes & \nodata \\
GW190926\_050336 & 3280$^{+3400}_{-1730}$ & 0.55$^{+0.44}_{-0.26}$ & 41.1 & 20.4 & BBH & 0.54 & 15.96 & 3.10 & 0.68 & No & 0.75 \\
GW190929\_012149 & 3130$^{+2510}_{-1370}$ & 0.53$^{+0.33}_{-0.20}$ & 66.3 & 26.8 & BBH & 0.87 & 32.58 & 6.06 & 1.36 & No & 0.55 \\
GW190930\_133541 & 770$^{+320}_{-320}$ & 0.16$^{+0.06}_{-0.06}$ & 14.2 & 6.9 & BBH & 1.00 & 12.34 & 3.26 & 0.51 & No & 0.95 \\
GW191103\_012549 & 990$^{+500}_{-470}$ & 0.20$^{+0.09}_{-0.09}$ & 11.8 & 7.9 & BBH & 0.94 & 32.74 & 8.34 & 1.81 & No & 0.75 \\
GW191105\_143521 & 1150$^{+430}_{-480}$ & 0.23$^{+0.07}_{-0.09}$ & 10.7 & 7.7 & BBH & 0.99 & 13.62 & 3.47 & 0.76 & No & 0.25 \\
GW191109\_010717 & 1290$^{+1130}_{-650}$ & 0.25$^{+0.18}_{-0.12}$ & 65.0 & 47.0 & BBH & 0.99 & 38.43 & 10.05 & 2.19 & No & 0.62 \\
GW191113\_071753 & 1370$^{+1150}_{-620}$ & 0.26$^{+0.18}_{-0.11}$ & 29.0 & 5.9 & BBH & 0.68 & 29.13 & 5.55 & 1.60 & No & 0.42 \\
GW191126\_115259 & 1620$^{+740}_{-740}$ & 0.30$^{+0.12}_{-0.13}$ & 12.1 & 8.3 & BBH & 0.70 & 15.14 & 3.77 & 0.84 & No & 0.79 \\
GW191127\_050227 & 3400$^{+3100}_{-1900}$ & 0.57$^{+0.40}_{-0.29}$ & 53.0 & 24.0 & BBH & 0.49 & 40.41 & 7.52 & 1.66 & No & 0.99 \\
GW191129\_134029 & 790$^{+260}_{-330}$ & 0.16$^{+0.05}_{-0.06}$ & 10.7 & 6.7 & BBH & 0.99 & \nodata & \nodata & \nodata & Yes & \nodata \\
GW191204\_110529 & 1800$^{+1700}_{-1100}$ & 0.34$^{+0.25}_{-0.18}$ & 27.3 & 19.3 & BBH & 0.74 & 70.74 & 12.82 & 2.83 & No & 0.77 \\
GW191204\_171526 & 650$^{+190}_{-250}$ & 0.13$^{+0.04}_{-0.05}$ & 11.9 & 8.2 & BBH & 0.99 & \nodata & \nodata & \nodata & Yes & \nodata \\
GW191215\_223052 & 1930$^{+890}_{-860}$ & 0.35$^{+0.13}_{-0.14}$ & 24.9 & 18.1 & BBH & 0.99 & 13.50 & 3.43 & 0.74 & No & 0.85 \\
GW191216\_213338 & 340$^{+120}_{-130}$ & 0.07$^{+0.02}_{-0.03}$ & 12.1 & 7.7 & BBH & 0.99 & 89.91 & 23.00 & 4.99 & No & 1.00 \\
GW191222\_033537 & 3000$^{+1700}_{-1700}$ & 0.51$^{+0.23}_{-0.26}$ & 45.1 & 34.7 & BBH & 0.99 & 16.87 & 4.22 & 0.92 & No & 0.89 \\
GW191230\_180458 & 4300$^{+2100}_{-1900}$ & 0.69$^{+0.26}_{-0.27}$ & 49.4 & 37.0 & BBH & 0.95 & 27.34 & 7.03 & 1.49 & No & 0.17 \\
GW200112\_155838 & 1250$^{+430}_{-460}$ & 0.24$^{+0.07}_{-0.08}$ & 35.6 & 28.3 & BBH & 0.99 & 54.39 & 10.51 & 2.22 & No & 0.82 \\
GW200128\_022011 & 3400$^{+2100}_{-1800}$ & 0.56$^{+0.28}_{-0.28}$ & 42.2 & 32.6 & BBH & 0.99 & 19.77 & 3.67 & 0.85 & No & 0.64 \\
GW200129\_065458 & 900$^{+290}_{-380}$ & 0.18$^{+0.05}_{-0.07}$ & 34.5 & 28.9 & BBH & 0.99 & 29.94 & 7.55 & 1.67 & No & 0.00 \\
GW200202\_154313 & 410$^{+150}_{-160}$ & 0.09$^{+0.03}_{-0.03}$ & 10.1 & 7.3 & BBH & 0.99 & 63.97 & 17.05 & 3.65 & No & 0.00 \\
GW200208\_130117 & 2230$^{+1000}_{-850}$ & 0.40$^{+0.15}_{-0.14}$ & 37.8 & 27.4 & BBH & 0.99 & \nodata & \nodata & \nodata & Yes & \nodata \\
GW200208\_222617 & 4100$^{+4400}_{-1900}$ & 0.66$^{+0.54}_{-0.28}$ & 51.0 & 12.3 & BBH & 0.70 & 27.76 & 5.63 & 1.20 & No & 0.30 \\
GW200209\_085452 & 3400$^{+1900}_{-1800}$ & 0.57$^{+0.25}_{-0.26}$ & 35.6 & 27.1 & BBH & 0.95 & \nodata & \nodata & \nodata & No & \nodata \\
GW200216\_220804 & 3800$^{+3000}_{-2000}$ & 0.63$^{+0.37}_{-0.29}$ & 51.0 & 30.0 & BBH & 0.77 & 34.58 & 9.41 & 1.46 & No & 1.00 \\
GW200219\_094415 & 3400$^{+1700}_{-1500}$ & 0.57$^{+0.22}_{-0.22}$ & 37.5 & 27.9 & BBH & 0.99 & 16.62 & 3.27 & 0.99 & No & 0.35 \\
GW200220\_061928 & 6000$^{+4800}_{-3100}$ & 0.90$^{+0.55}_{-0.40}$ & 87.0 & 61.0 & BBH & 0.62 & 18.08 & 4.97 & 1.04 & No & 0.34 \\
GW200220\_124850 & 4000$^{+2800}_{-2200}$ & 0.66$^{+0.36}_{-0.31}$ & 38.9 & 27.9 & BBH & 0.20 & 54.52 & 14.94 & 3.14 & No & 0.59 \\
GW200224\_222234 & 1710$^{+490}_{-640}$ & 0.32$^{+0.08}_{-0.11}$ & 40.0 & 32.5 & BBH & 0.99 & 17.68 & 3.34 & 0.72 & No & 0.00 \\
GW200225\_060421 & 1150$^{+510}_{-530}$ & 0.22$^{+0.09}_{-0.10}$ & 19.3 & 14.0 & BBH & 0.99 & 42.91 & 8.18 & 1.80 & No & 0.28 \\
GW200302\_015811 & 1480$^{+1020}_{-700}$ & 0.28$^{+0.16}_{-0.12}$ & 37.8 & 20.0 & BBH & 0.91 & 35.62 & 9.17 & 2.08 & No & 0.78 \\
GW200306\_093714 & 2100$^{+1700}_{-1100}$ & 0.38$^{+0.24}_{-0.18}$ & 28.3 & 14.8 & BBH & 0.24 & 31.47 & 7.96 & 1.70 & No & 0.37 \\
GW200308\_173609 & 5400$^{+2700}_{-2600}$ & 0.83$^{+0.32}_{-0.35}$ & 36.4 & 13.8 & BBH & 0.86 & 21.93 & 5.57 & 0.91 & No & 0.69 \\
GW200311\_115853 & 1170$^{+280}_{-400}$ & 0.23$^{+0.05}_{-0.07}$ & 34.2 & 27.7 & BBH & 0.99 & 60.51 & 14.76 & 3.28 & No & 1.00 \\
GW200316\_215756 & 1120$^{+470}_{-440}$ & 0.22$^{+0.08}_{-0.08}$ & 13.1 & 7.8 & BBH & 0.99 & \nodata & \nodata & \nodata & Yes & \nodata \\
GW200322\_091133 & 3600$^{+7000}_{-2000}$ & 0.60$^{+0.84}_{-0.30}$ & 34.0 & 14.0 & BBH & 0.08 & 47.76 & 12.39 & 2.88 & No & 0.73 \\
\enddata
\end{deluxetable*}

\end{document}